\documentclass[twocolumn,prx,showpacs,amsmath,superscriptaddress]{revtex4-1}
\usepackage{epsfig,amsfonts,amsbsy,yfonts}
\usepackage{color}
\usepackage{verbatim}
\usepackage{hyperref}
\hypersetup{
  colorlinks   = true, 
  urlcolor     = blue, 
  linkcolor    = blue, 
  citecolor   = blue 
}

\graphicspath{{./Figures/}}

\newcommand{\state}[1]{\left \vert \left. #1 \right\rangle   \right.}
\newcommand{\bstate}[1]{\left\langle  \left.  #1  \right \vert \right. }

\newcommand{\up}{\uparrow}
\newcommand{\dw}{\downarrow}

\newcommand{\unitvector}{\vec{\mathsf n }_\lambda}

\def\be{\begin{equation}}
\def\ee{\end{equation}}
\def\beq{\begin{eqnarray}}
\def\eeq{\end{eqnarray}}
\def\ba#1{\begin{array}{#1}}
\def\ea{\end{array}}
\def\bn{\begin{enumerate}}
\def\en{\end{enumerate}}

\usepackage{amsthm}
\usepackage{mathtools}
\theoremstyle{remark}

\newcommand{\bea}{\begin{eqnarray}}
\newcommand{\eea}{\end{eqnarray}}
\newcommand{\beas}{\begin{eqnarray*}}
\newcommand{\eeas}{\end{eqnarray*}}

\newcommand{\bquo}{\begin{quote}}
\newcommand{\enqu}{\end{quote}}
\renewcommand{\(}{\begin{equation}}
\renewcommand{\)}{\end{equation}}

\def\Tr{ \hbox{\rm Tr}}

\def\Tr{ \hbox{\rm Tr}}

\begin{document}
\title{Adiabatic landscape and optimal paths in ergodic systems}

\author{Sho Sugiura}
\affiliation {Physics and Informatics Laboratories, NTT Research, Inc., East Palo Alto CA, United States}
\affiliation{Department of Physics, Harvard University, Cambridge MA, United States
}

\author{Pieter W. Claeys}
\affiliation {Department of Physics, Boston University, Boston MA, United States}
\affiliation{TCM Group, Cavendish Laboratory, University of Cambridge, Cambridge, UK}
\author{Anatoly Dymarsky}
\affiliation {Department of Physics, University of Kentucky, Lexington KY, United States}
\affiliation{Skolkovo Institute of Science and Technology, Skolkovo Innovation Center, Moscow, Russia}
\author{Anatoli Polkovnikov}
\affiliation {Department of Physics, Boston University, Boston MA, United States}

\date{\today}

\begin{abstract}
Whether one is interested in quantum state preparation or in the design of efficient heat engines, adiabatic (reversible) transformations play a pivotal role in minimizing computational complexity and energy losses. Understanding the structure of these transformations and identifying the systems for which such transformations can be performed efficiently and quickly is therefore of primary importance. In this paper we focus on finding optimal paths in the space of couplings controlling the system's Hamiltonian. More specifically, starting from a local Hamiltonian we analyze directions in the space of couplings along which  adiabatic transformations can be accurately generated by local operators, which are both realizable in experiments and easy to simulate numerically. We consider a non-integrable 1D Ising model parametrized by two independent couplings, corresponding to longitudinal and transverse magnetic fields. We find regions in the space of couplings characterized by a very strong anisotropy of the variational adiabatic gauge potential (AGP), generating the adiabatic transformations, which allows us to define optimal adiabatic paths. We find that these paths generally terminate at singular points characterized by extensive degeneracies in the energy spectrum, splitting the parameter space into adiabatically disconnected regions. The anisotropy follows from singularities in the AGP, and we identify special robust weakly-thermalizing and non-absorbing many-body ``dark'' states which are annihilated by the singular part of the AGP and show that their existence extends deep into the ergodic regime.
\end{abstract}

\maketitle

\section{Introduction}

With the rapid progress of quantum technologies, the design of efficient protocols to control and numerical methods to describe quantum systems quickly moved to the forefront of current research. To achieve a better performance, a crucial element is the ability to perform adiabatic transformations, i.e. transformations between states that are adiabatically connected \cite{kolodrubetz_geometry_2017}. 
For example, quantum annealing and adiabatic quantum computation are based on an adiabatic process transforming a simple initial ground state into a final non-trivial eigenstate, and were shown to be a universal tool for quantum computation \cite{albash_adiabatic_2018}. Likewise, any quantum gate operation can be designed using an adiabatic protocol \cite{aharonov_adiabatic_2007,nielsen_chuang_2010}. The experimental preparation of equilibrium states in isolated or nearly-isolated systems such as cold atoms or NV centers is often achieved by adiabatic transformations of the Hamiltonian, starting from a simple initial state. In some cases, including Floquet-engineered systems \cite{bukov_universal_2015,dalessio_long-time_2014,goldman_periodically_2014}, such a procedure is not only convenient but is actually required, since these systems do not naturally thermalize by interacting with their environment. In the context of thermodynamics, adiabatic (reversible) processes are also of crucial importance. They allow one to minimize the dissipative losses associated with an increase of entropy and achieve the maximal possible efficiency of  energy conversion, e.g.~in heat engines and refrigerators \cite{jarzynski_equalities_2011,campo_more_2014}. On the theoretical side, adiabatic transformations underly many concepts including the Schrieffer-Wolff transformation \cite{schrieffer_relation_1966,bravyi_schriefferwolff_2011,wurtz_variational_2020}, 
and the dressing of quasiparticles by interactions underlying e.g. Fermi liquid theory \cite{landau_statistical_1980}. Such transformations not only allow us to theoretically understand the properties of low-energy Hamiltonians, but also provide a convenient tool to greatly improve the efficiency of numerical methods, allowing one to focus on particular subspaces of interest \cite{wurtz_variational_2020}. 

A standard limitation of our ability to use adiabatic transformations is that they, almost by definition, have to be extremely slow. In many-body interacting systems, unless we are  interested in the ground state of a gapped system, the necessary time scales are exponentially large with the system size \cite{jarzynski_geometric_1995,kolodrubetz_geometry_2017}. A similar exponential slowing down is required if we are interested in following a ground state which either crosses a first-order phase transition \cite{de_grandi_adiabatic_2010,del_campo_assisted_2012}, enters a quantum glass regime \cite{mezard_spin_1987,nishimori_statistical_2001}, or follows from an annealing protocol solving a hard computational problem~\cite{farhi_science_2001}. 

From the computational point of view, strict upper bounds on the rate of parameter change result in heavy numerical costs. On the experimental side, they lead to a very slow state preparation and large energy processing times. Moreover, the necessary long time scales are generally inaccessible in experimental setups. Systems cannot be perfectly isolated from their environment, leading to decoherence and noise which can destroy the state or erase the information that adiabaticity is trying to preserve. Rather recently, it was realized that this problem can be circumvented and adiabatic transformations can be sped up, in principle arbitrarily, by adding an additional term to the Hamiltonian, suppressing all dynamical/diabatic transitions. Such ideas were first introduced in 2003 by M. Demirplak and S. Rice \cite{demirplak_adiabatic_2003} and independently in 2009 by M. Berry \cite{berry_transitionless_2009} and were subsequently termed counterdiabatic (CD) or transitionless driving. The topic of counterdiabatic driving and the related field of shortcuts to adiabaticity has recently gained tremendous attention in both experimental and theoretical literature \cite{del_campo_shortcuts_2013,guery-odelin_shortcuts_2019,torrontegui_chapter_2013,baksic_speeding_2016,claeys_floquet-engineering_2019,petiziol_accelerating_2019,petiziol_fast_2018,theis_counteracting_2018,vepsalainen_optimal_2018,zhou_floquet-engineered_2019,del_campo_focus_2019}. 

In counterdiabatic protocols one applies an additional term to the Hamiltonian, proportional to the generator of adiabatic transformations, the so-called adiabatic gauge potential (AGP). This extra term suppresses all diabatic(non-adiabatic) excitations/losses. The main difficulty of this approach is that the AGP is generally highly non-local. Furthermore, the AGP is not only useful in counterdiabatic driving, but also contains a wealth of information on the geometry of eigenstates and diabatic response \cite{kolodrubetz_geometry_2017} and serves as a very sensitive probe of quantum chaos~\cite{mohit_2020}. The exact AGP is local only in some special situations, including symmetry transformations or transformations of the ground state of a gapped system \cite{hastings_quasiadiabatic_2005,bravyi_short_2011,bachmann_automorphic_2012}. Fortunately, even if the exact AGP is generally out of reach, it was recently realized that in some specific instances we can find an approximate yet accurate local AGP using a variational minimization \cite{sels_minimizing_2017,kolodrubetz_geometry_2017,claeys_floquet-engineering_2019}. The resulting local AGP was shown to be highly efficient both in solving computationally difficult problems~\cite{Hartmann_2019, passarelli_2020} and performing efficient Schrieffer-Wolff transformations \cite{wurtz_variational_2020, wurtz_emergent_2020}. Still, several general and unanswered questions remain: \emph{(i) When do such local approximations apply? (ii) Which are the optimal protocols for local adiabatic evolution? (iii) Can we learn which states are most heavily affected by diabatic effects from these local approximations and is it possible to identify the states for which dissipation is minimal?}

In this work, we first focus on finding an optimal path in the space of system's parameters to design local protocols for adiabatic state preparation. Very often, physical systems are controlled by multiple parameters, e.g. pressure, temperature, chemical potential, external electric and magnetic fields in thermodynamics or single-spin controls and two-spin interactions in quantum control. While the order in which these parameters are changed will not matter if everything happens perfectly adiabatically, the diabatic effects can vary drastically depending on how these parameters are tuned. It is then natural to ask for the optimal path in the space of parameters, minimizing diabatic effects. This will be the focus of this work.

For concreteness, we will consider protocols satisfying the time-dependent Schr\"odinger equation (we set $\hbar=1$ throughout the text),
\begin{equation}
i\frac{\partial}{\partial t} \left| \psi(t) \right\rangle = H(\vec \lambda(t))\left| \psi(t) \right\rangle, \label{time dep eq}
\end{equation}
where the Hamiltonian depends on a set of time-dependent control parameters $\vec{\lambda}(t)$ and we initialize the system at $t=0$ in a stationary eigenstate of $H(\vec \lambda(0))$ (all our results immediately extend to mixed initial states). The question is then how to vary $\vec{\lambda}(t)$ such that the state remains close to an instantaneous eigenstate of $H(\vec{\lambda(t)})$. To answer this question, we analyze the adiabatic landscape of  a fairly generic non-integrable 1D Ising model characterized by two independent couplings (cf. Eq.~\eqref{TFI}). Specifically, we show that the variational adiabatic gauge potential (VAGP), which gives the best local approximation to the exact AGP (see Eqs.~\eqref{eq:def_eig} and \eqref{eq:def_AGP}), forms a two-dimensional vector space, and the directions where the norm of the VAGP is minimal define the optimal paths minimizing diabatic effects. We mainly focus on infinite temperature states, where the equilibrium properties of the system are completely featureless. Nevertheless, the problem of adiabatic continuation remains well defined and highly nontrivial. We find that the evolution along the optimal direction is efficient; that is, eigenstates which are drawn from the middle of the spectrum remain close to the instantaneous eigenstates, maintaining small energy variance. As we will show below (see also Refs.~\cite{claeys_floquet-engineering_2019,mohit_2020}) the AGP can be expressed through the long-time limit of non-equal time correlation functions of the operators conjugate to the coupling. Therefore, they cannot be analyzed by the methods of equilibrium statistical mechanics. Our findings thus imply that temperature plays a much smaller role in adiabatic transformations than in equilibrium settings. 

\begin{figure}[ht!]
    \includegraphics[width=1.\columnwidth]{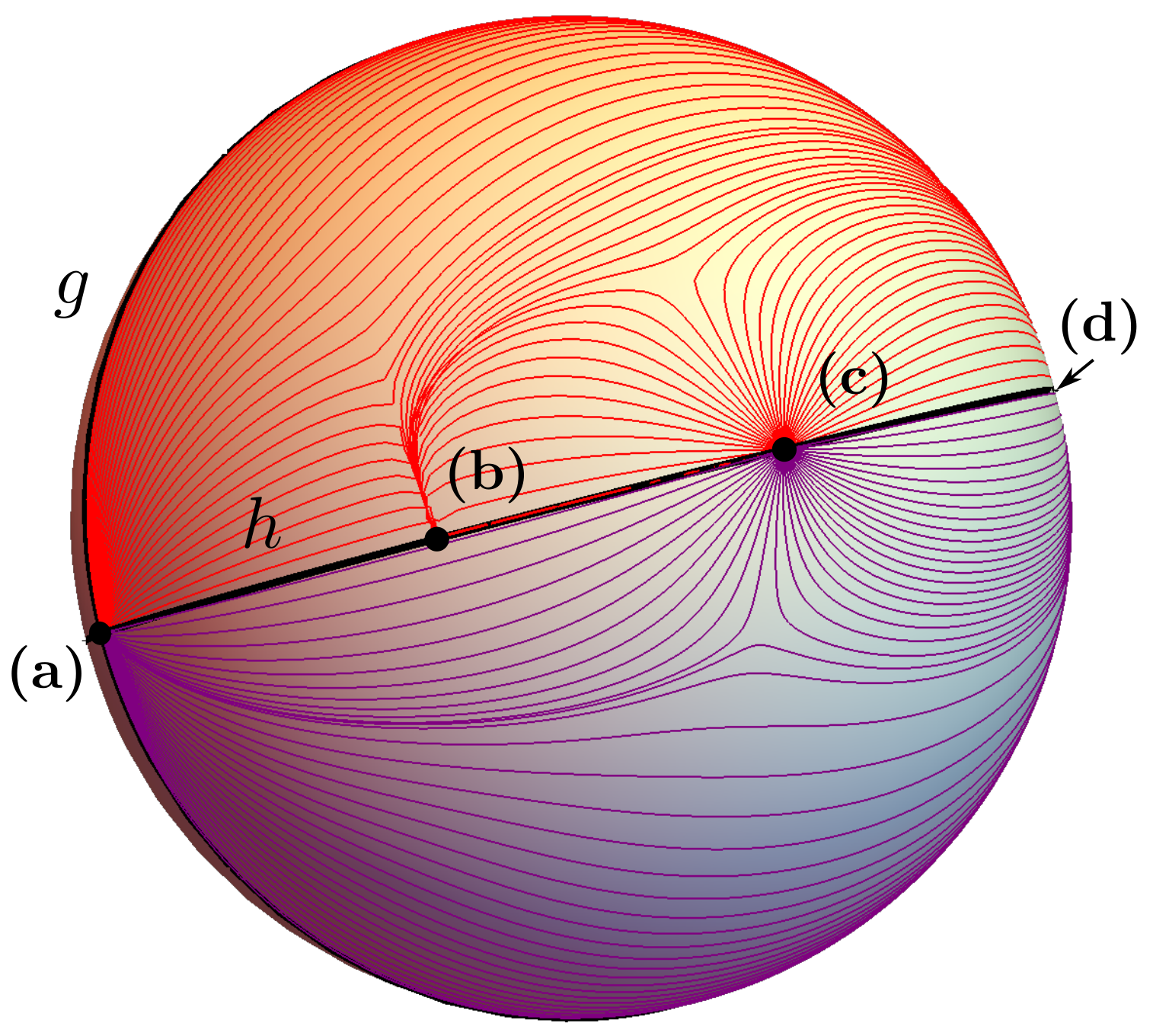}
    \caption{Flow diagram indicating the optimal path for quantum control using a 3-body (purple) and 5-body ansatz (red) for a two-dimensional parameter space $(h,g)$. The horizontal and vertical axes are $h$ and $g$, respectively, with the two poles of the sphere given by $(h,g)=(0,0)$ [(a)] and $(\infty,\infty)$ [(d)]. Source flows can be observed at $(h,g)=(0,0)$ [(a)] and $(2,0)$ [(c)], where the optimal direction is approximately the radial one. The norm of the variational gauge potential is highly anisotropic: near the source flows the norm is small and nearly system-size independent along the optimal directions, while increasing drastically in the orthogonal direction and diverging exactly at the points (a) and (c). For the 5-body ansatz an additional singular point appears at $(h,g)=(1,0)$ [(b)], strongly disrupting the optimal directions in its vicinity.}
    \label{Flow Diagram sphere}
\end{figure}

Let us now introduce the Hamiltonian that we will analyze in this work, describing the quantum Ising model in the presence of a longitudinal and transverse field, as
\begin{align}
	H = J \sum_{i} \sigma_i^z\sigma_{i+1}^z + h \sum_{i} \sigma_i^z + g \sum_i \sigma_i^x,
\end{align}
and we introduce a shorthand notation that is convenient for translationally-invariant systems
\begin{align}
	H = J {Z}{Z} + h {Z} + g {X}	\label{TFI},
\end{align}
with
\[
ZZ\equiv \sum_{i=1}^L\sigma^z_i\sigma^z_{i+1},\quad
Z \equiv \sum_{i=1}^L\sigma^z_i,
\quad X\equiv \sum_{i=1}^L\sigma^x_i, 
\]
and so on. Fixing the coupling in front of the Ising interaction $ZZ$ to be unity, $J=1$, $h$ and $g$ will be taken as control parameters throughout this paper.  The main results of our paper are summarized in Fig.~\ref{Flow Diagram sphere}: each point represents a choice of couplings defining a Hamiltonian, and the lines show the optimal adiabatic directions presented as a flow diagram. As will be discussed later, this diagram has a very rich structure and is in many respects similar to the standard equilibrium phase diagrams (except that, as already pointed out, it corresponds to an infinite temperature).  Let us now summarize the most essential findings reflected in this figure, which will be explained in detail in the paper.

\begin{itemize}

 \item  Along the $h$-axis the flow diagram contains singularities, corresponding to Hamiltonians with exponentially-large degeneracies in the energy spectrum, which we call macroscopic degeneracies. These singularities serve as sources/sinks of adiabatic flows and play a similar role to critical points in equilibrium phase diagrams.
 
\item Close to these singularities, the VAGP becomes infinitely anisotropic, with highly-anisotropic regions extending far away from the singularities. This high anisotropy implies that optimal directions, along which local adiabatic transformations are highly efficient, remain well defined. Such optimal directions define paths with minimal dissipation and maximum fidelity for state preparation.

\item Near the singular points, there are special many-body ``dark'' states annihilated by the diverging part of the VAGP. These states exist throughout the entire energy spectrum; similar to the anisotropic regions these are highly robust and extend deep into the ergodic regimes, bearing many parallels with the recently-discovered quantum scars and the eigenstates of constraint models \cite{bernien_probing_2017,turner_weak_2018,turner_quantum_2018,moudgalya_exact_2018,sugiura_many-body_2019,khemani_signatures_2019,james_nonthermal_2019,ho_periodic_2019,choi_emergent_2019}. While these states form an exponentially-small fraction of the total Hilbert space, their total number can still be exponentially large; as they are immune to the usual dissipation they can be efficiently prepared both numerically and experimentally.

\item The optimal adiabatic directions allow us to define adiabatic flows similar to the renormalization group flows, as shown in the figure, and these flows in turn define adiabatically-connected families of Hamiltonians. The norm of the exact AGP is equivalent to the Fubini-Study metric defining the distance between eigenstates of adiabatically-connected Hamiltonians \cite{kolodrubetz_geometry_2017}. Therefore, these flows can be interpreted as lines approximately minimizing the local distance between eigenstates (or more accurately between energy shells) of different Hamiltonians. Along these flows, both states and operators can be dressed to a very good accuracy under the unitary transformations generated by the local VAGP. In particular, such directions are characterized by the existence of nearly-conserved operators, which are locally-dressed operators conjugate to the coupling along these directions.

\item Near the singular points, the VAGP diverges in all directions except for the optimal one. However, the divergent part of the VAGP is a well-defined local operator, implying that a local dressing can be used to efficiently perform adiabatic rotations near these singularities. Combined with the fact that all adiabatic flows terminate at one of such singularities, we arrive at the interesting conclusion that any optimal adiabatic path between two generic points goes through one of these singularities. In other words, the system first has to be brought to the singular point, then a local rotation needs to be performed, before going to the target point along a different flow line. Importantly, such a path can be always found locally by following the optimal direction of the adiabatic flow. 

\item The optimal directions generally depend on the support/size of the variational ansatz (see the top and bottom halves in Fig.~\ref{Flow Diagram sphere}), i.e. the support of the operator generating approximate adiabatic  transformations. New singularities appear in the higher-order variational ansatz with an increased local support, reflecting higher-order divergences in the perturbative expansion of the AGP. These singularities arise from the degeneracies associated with higher-order interactions and appear at rational couplings, bearing many similarities to the divergences appearing in both KAM theory~\cite{Wayne_notes} and locator expansions~\cite{scardicchio_2017}. The emergence of higher-order singularities indicates that it is not possible to improve local dressing, by either adding additional local terms to the CD protocol or by slowing down the ramping rate in the absence of CD driving, without abruptly altering the path near these new singularities.

\item The adiabatic flow diagram remains well-defined even at infinite temperature, where no structure exists in the equilibrium state according to statistical mechanics. Interestingly, many of its features persist at all temperatures, all the way down to the ground state at zero temperature.

\end{itemize}

We confirm these general findings with numerical simulations for the non-integrable 1D Ising  model described by the Hamiltonian~\eqref{TFI}. Our results can have a broad range of applications in various problems, beyond simply finding optimal paths for annealing or state preparation. In particular, they can be used to find efficient local conservation laws and corresponding ``most-integrable'' directions, to find the nearest integrable (simple) points that are locally connected to a Hamiltonian of interest, to define most efficient ways of obtaining effective low-energy theories starting from a noninteracting model, and so on.

This paper is organized as follows: In Sec.~\ref{sec:VAGP}, we introduce the VAGP and define the optimal adiabatic directions. Applications to approximate CD driving and slowest operators are also explained there. Sec.~\ref{sec:flow diagram} is the highlight of this paper, where we obtain the flow diagram that defines the optimal directions at each point of the coupling space. We demonstrate that both for conventional adiabatic driving and for the approximate CD protocols state preparation along the optimal paths shows a much better performance than along the orthogonal directions. We explain that the flows terminate/start at special sources/sinks, where the VAGP develops divergencies in the orthogonal directions, becoming infinitely anisotropic, and show how these singularities arise from the perturbative expansion of the exact AGP. We then explain the emergence of special dark states unaffected by the singular part of the VAGP.  In Sec.~\ref{sec:bigger ansatz}, we study how the VAGP depends on the size of the variational ansatz and explain the emergence of new singularities near rational values of $h$. We then use the VAGP to construct approximate local conserved operators and analyze their life times in Sec.~\ref{sec: conserved operator}. Details of the perturbative expansion are given in Sec.~\ref{sec:perturbative} and Sec.~\ref{sec:conclusion} is reserved for conclusions.

\section{Variational Adiabatic Gauge Potential}\label{sec:VAGP}

In this section we will give a brief introduction to the concept of the (variational) adiabatic gauge potential, emphasizing its structure as a vector in a system with multiple controls (tunable parameters). Much of this discussion can be found in earlier papers \cite{sels_minimizing_2017,kolodrubetz_geometry_2017,claeys_floquet-engineering_2019}, but is included here in order to be self-contained and to make an explicit connection of VAGP with slow operators \cite{kim_slowest_2015,michailidis_slow_2018}, operator spreading \cite{nahum_operator_2018,khemani_operator_2018,von_keyserlingk_operator_2018,gopalakrishnan_hydrodynamics_2018,swingle_unscrambling_2018,parker_universal_2019,avdoshkin2019euclidean}, and emergent conservation laws \cite{mierzejewski_approximate_2015}, which will be relevant for the presented flow diagram.

\subsection{Theoretical background}

Let us consider a family of Hamiltonians $H(\vec \lambda)$, where $\vec \lambda$ specifies the space of available couplings or controls. Any protocol corresponds to a time-dependent choice of $\vec{\lambda}(t)$, with an adiabatic protocol corresponding to a vanishing time-derivative $|\dot{\vec{\lambda}}(t)|$. 

The effects of time-dependent couplings are most clearly illustrated in the instantaneous (co-moving) eigenstates of the Hamiltonian $|n(\vec \lambda)\rangle$, satisfying
\be\label{eq:def_eig}
H(\vec \lambda)|n(\vec \lambda)\rangle = \epsilon_n(\vec \lambda) |n(\vec \lambda)\rangle.
\ee
Any change in the control parameters corresponds to a change in the eigenstates, and one can formally define the adiabatic gauge potential (AGP) as the Hermitian operator $\vec{\mathcal{A}}(\vec \lambda)$ generating these basis changes \cite{kolodrubetz_geometry_2017}:
\be\label{eq:def_AGP}
i \partial_{j} |n(\vec \lambda)\rangle=\mathcal{A}_j (\vec \lambda) |n(\vec \lambda)\rangle,
\ee
in which $\partial_{j}$ is the partial derivative w.r.t. $\lambda_j$. Note that, since eigenstates are only defined up to a phase (or more general rotations in the presence of degeneracies), the AGP is not uniquely defined and supports a gauge freedom. 

We will be interested in finding the time evolution \eqref{time dep eq} of an initial pure state $|\psi(t=0)\rangle$ under time evolution governed by a time-dependent Hamiltonian $H(\vec{\lambda}(t))$, where the only explicit time dependence is through the control parameters~\footnote{Our discussion equally applies to the evolution of mixed states}.  Expanding this state in the co-moving basis 
\be
|\psi(t)\rangle=\sum_n a_n(t) |n(\vec \lambda(t))\rangle,
\ee
it is easy to check that the time evolution in this new basis is governed by the moving Hamiltonian
\begin{align}\label{eq:mfHam}
H_m(t)& = H(\vec \lambda(t))-\sum_j \dot \lambda_j \mathcal{A}_j (\vec \lambda (t)), \nonumber\\
& = H(\vec{\lambda}(t))-\dot{\vec{\lambda}}(t) \cdot \vec{\mathcal{A}}(\vec{\lambda}(t)).
\end{align}
Specifically, \
\begin{align}
\label{comoving}
& i \dot{a}_n(t) =\sum_{l} H_m^{n l}(t)\,  a_l (t),\\
&H_m^{nl}(t)= \langle n(\vec{\lambda}(t))|H_m(t)|l(\vec{\lambda}(t))\rangle, \nonumber
\end{align}
which takes the form of a regular matrix representation of the Schr\"odinger equation, but with time-dependent basis states, which are accounted for by the second term in Eq.~\eqref{eq:mfHam}. In the limit $\dot{\vec{\lambda}} \to 0$  this additional term vanishes such that there are no transitions between instantaneous eigenstates of $H(\vec{\lambda})$. At non-vanishing $|\dot{\vec{\lambda}}|$ the extra term in the moving Hamiltonian, proportional to the AGP, cannot be neglected. Since $H(\vec{\lambda})$ is by construction diagonal in the co-moving frame, all diabatic excitations/losses are generated by the off-diagonal elements of the AGP.

Following Ref.~\cite{kolodrubetz_geometry_2017}, Eq.~\eqref{eq:def_AGP} can be recast as an operator equation
\begin{align}
	\left[H, G_j(\vec{\mathcal{A}}) \right] = 0, 
	\label{equation exact gp}
\end{align}
in which
\begin{align}
	G_j(\vec{\mathcal{A}}) \equiv \partial_{j} H + {i }[\mathcal{A}_j,H].
\label{eq:G_def}
\end{align} 
The matrix $G_j(\vec{\mathcal{A}})$ is diagonal in the eigenbasis of $H$ and its diagonal matrix elements are given by $\partial_j \epsilon_n(\vec \lambda)$, the generalized forces conjugate to $\lambda_j$. In other words, one can view any infinitesimal deformation of the Hamiltonian along the $\lambda_j$ direction $\partial_j H$ as consisting of a spectrum change encoded in $G_j$ and an eigenbasis rotation encoded in $\mathcal{A}_j$. 

Eq.~\eqref{equation exact gp} remains well-defined in both the classical and thermodynamic limits. However, with the exception of symmetry transformations/integrable systems, the solutions to this equation are generally unstable to infinitesimal perturbations and might not even exist in either of these limits~\cite{jarzynski_geometric_1995,kolodrubetz_geometry_2017, mohit_2020}. Therefore, finding approximate local gauge potentials is essential to circumvent this problem. One goal of this paper is to convey that, even though the exact AGP might be ill-defined, such local approximations can be well-defined and meaningful.

A particularly powerful approach to finding approximate solutions is the variational method. It is based on the observation  that Eq.~\eqref{equation exact gp} can be interpreted as the  minimization condition for  the auxiliary action $S(\vec{\mathcal A})$ \cite{sels_minimizing_2017}
 \begin{align}
	{\delta S \over \delta \mathcal{A}_j} = 0, \quad \textrm{with}\quad 	S\equiv \sum_j {\rm Tr}[G_j(\vec{\mathcal A})^\dagger G_j(\vec{\mathcal A})].
	\label{equation variational}
\end{align}
Approximate solutions of Eq.~\eqref{equation exact gp} can be found by choosing a specific subset of operators as an ansatz for the AGP and finding the minimum of the action. We call the resulting solution the (local) variational adiabatic gauge potential (VAGP). Also note that the action for the VAGP in Eq.~\eqref{equation variational} can be interpreted as the action at infinite temperature. In principle, it can be extended to finite temperatures through the introduction of a thermal state $\exp\left[-\beta H\right]$ in $S$ (see Ref.~\cite{sels_minimizing_2017}), although this strongly complicates the resulting minimization. In this paper we focus on variational manifolds consisting of all local operators with a given support (see Sec. \ref{subsec:opt_adiabatic_dir} for details). One can develop a similar expansion based on nested commutators of $\partial_{j} H$ and $H$ \cite{claeys_floquet-engineering_2019}. We checked that this second expansion leads to very similar conclusions. Despite being an approximate solution, as we discuss below, the local VAGP can be used to determine highly nontrivial properties of the system. Let us mention a few of them.

\emph{Approximate counterdiabatic driving}. -- The notion of counterdiabatic (CD) driving immediately follows from this derivation, since the exact solution of $\vec{\mathcal{A}}$ can be used to completely suppress energy dissipation by evolving a system with the CD Hamiltonian including an additional term $\dot{\vec{\lambda}}\cdot\vec{\mathcal{A}}(\vec{\lambda})$,
\begin{align}
	H_{\rm CD}(t) = H(\vec{\lambda}(t))+\dot{\vec{\lambda}}(t)\cdot\vec{\mathcal{A}}(\vec{\lambda}(t)). \label{Ham CDD}
\end{align}
Representing the evolution in the co-moving frame of $H(\vec{\lambda}(t))$, the additional counterdiabatic term cancels, such that the moving frame Hamiltonian is exactly given by $H(\vec{\lambda}(t))$, which is diagonal and hence does not lead to any excitations or dissipation. Namely, starting from any energy eigenstate $|\psi(t=0)\rangle = |n(\vec{\lambda}(0))\rangle$ the state at later times remains an instantaneous eigenstate $|n(\vec{\lambda}(t))\rangle$. In the limit of an infinitely fast rate of change $|\dot {\vec \lambda}|\to\infty$, the AGP dominates, and the resulting evolution can be seen as a pure dressing of the initial state.  We will refer to a protocol corresponding to $ H(\vec{\lambda}(t))$, where no CD term is present, as the unassisted protocol.

While the exact AGP generally cannot be realized in many-body systems, the use of local approximations from the variational minimization has already been shown to lead to a significant suppression of transitions \cite{sels_minimizing_2017,claeys_floquet-engineering_2019, Hartmann_2019, tamiro_2019, passarelli_2020}. As such, the availability of an accurate local VAGP can also be used to reduce dissipation and design efficient annealing protocols.

\emph{Approximate state dressing}. -- Starting from an initial eigenstate of the instantaneous Hamiltonian, counterdiabatic driving can be interpreted as interpolating between two limits: $\dot{\vec{\lambda}}\to 0$ returns adiabatic state preparation, whereas $\dot{\vec{\lambda}} \to \infty$  dresses the initial state with the (approximate) gauge potential. Namely, in this limit the Schr\"odinger equation reduces to
\begin{align}
i  \partial_t \left|\psi(t)\right\rangle = \dot{\vec{\lambda}}(t)\cdot\vec{\mathcal{A}}(\vec{\lambda}(t))\left|\psi(t)\right\rangle
\end{align}
For an exact AGP, $\left|\psi(t)\right\rangle = |\psi(\vec{\lambda}(t))\rangle$ and this equation reduces to
\begin{align}
i  \partial_{\vec{\lambda}} |\psi(\vec{\lambda})\rangle = \vec{\mathcal{A}}(\vec{\lambda})|\psi(\vec{\lambda})\rangle.
\end{align}
This corresponds to a (quasi-)adiabatic dressing of the initial state \cite{hastings_quasiadiabatic_2005,bachmann_automorphic_2012, wurtz_variational_2020, wurtz_emergent_2020}. The possibility of such dressing with a (quasi-)local $\mathcal{A}$ is a crucial ingredient in classifying topological phases, where all ground states within a given phase can be adiabatically connected using a local dressing.

\emph{Operator spreading.} -- A formal solution to Eq.~\eqref{equation exact gp} can be found using the Lehmann's representation as
\be
\mathcal{A}_j=-\lim_{\epsilon \to 0^+}{1 \over 2}\int_{-\infty}^{\infty} dt\  {\rm sgn}(t)\ \mathrm e^{-\epsilon |t|}\left( {\partial_{j}} H\right)(t),
\label{eq:A_Heisenberg}
\ee
where
\be
\left({\partial_{j}} H\right)(t)\equiv \mathrm e^{i H t} \left(\partial_{j} H \right) \mathrm e^{-i H t} 
\ee
is the operator conjugate to the parameter $\lambda_j$, $\partial_{j}H$, in the Heisenberg representation w.r.t. the instantaneous Hamiltonian $H$. For classical Hamiltonian systems, this representation originates from C. Jarzynski \cite{jarzynski_geometric_1995}. As mentioned before, the exact solution is highly sensitive to the choice of $\partial_{j}H$ and the limit $\epsilon \to 0$ will generally diverge in chaotic systems. Keeping $\epsilon$ finite then corresponds to finding an approximate AGP, which will be local for a local $\partial_{j}H$ due to the finite support of $\left({\partial_{j}} H\right)(t)$ at finite times, following recent results on operator spreading (e.g.~\cite{swingle_unscrambling_2018}) and Lieb-Robinson bounds \cite{Lieb:1972aa}. This representation has also been combined with the variational principle to find an efficient variational ansatz in chaotic many-body systems \cite{claeys_floquet-engineering_2019}.

\emph{Conservation laws and slowest operators.} -- \label{sec:conservation_laws} A local AGP immediately implies an additional local conservation law, since $G_j(\vec{\mathcal{A}})$ by definition commutes with the Hamiltonian. Minimizing the action then corresponds to obtaining a `slowest operator' \cite{kim_slowest_2015}, minimizing the commutator with the Hamiltonian (setting the time scale for thermalization), which then becomes an exact conserved quantity if the local VAGP becomes an exact AGP. Interestingly, if we consider the representation of the AGP through Eq.~\eqref{eq:A_Heisenberg} with finite $\epsilon$, the corresponding $G_j(\vec{\mathcal{A}})$ exactly coincides with the approximately-conserved operator obtained by the time-averaging of $\left({\partial_{j}} H\right)(t)$ introduced in Ref.~\cite{mierzejewski_approximate_2015}. In particular, using Eq.~\eqref{eq:A_Heisenberg} it is easy to check that 
\be
G_j(\vec{\mathcal{A}})= \overline {\left({\partial_{j}} H\right)}\equiv {\epsilon\over 2}\int_{-\infty}^{\infty} dt\ \mathrm e^{-\epsilon |t|} \left({\partial_{j}} H\right)(t),
\ee
namely, it is the part of $\partial_{j}H$ that is conserved and does not decay with time.

\subsection{Optimal adiabatic directions}
\label{subsec:opt_adiabatic_dir}
From Eq.~\eqref{eq:mfHam} it can be seen that all diabatic transitions are induced by the AGP. For a time-dependent change along a certain direction $\dot{\vec{\lambda}} = |\dot{\lambda}|\, \unitvector$, for a fixed rate of change $|\dot{\lambda}|$ along a direction set by a unit vector $\unitvector$, these transitions can be expected to be maximally suppressed along directions where the norm of $\mathcal{A}_{\lambda}= \unitvector \cdot \vec{\mathcal{A}}$ is minimal. In the same way that the gap between the ground state and the first excited state sets the time scale for quantum annealing, the norm of the AGP along a certain direction sets the scale for the rate of change of the control parameter $|\dot \lambda|$: for small $||\mathcal{A}_\lambda||$ the control parameter can be changed rather fast without inducing large diabatic effects, whereas for large $||\mathcal{A}_\lambda||$ even slow deformations of the Hamiltonian immediately lead to diabatic transitions. While the local VAGP is not exact, it contains information about transitions through local interactions, which are often the most damaging because they can lead to a large energy transfer. We will demonstrate below that this is indeed the case.

Given a multi-dimensional space of control parameters, one can thus set the optimal direction as the direction for which the norm of the VAGP is minimal. In principle, one can define different norms, so the minimization procedure is not unique. For example, one could choose norms tailored for particular states, e.g. the ground state. In this work we will use the Fr\"obenius (L2) trace-norm, equivalent to the common infinite-temperature norm. These norms have the advantage that they can be easily calculated in large systems (including the thermodynamic limit) without any need to diagonalize the Hamiltonian. As such, the actual minimization is particularly straightforward. Remarkably, it was observed in Refs. \cite{sels_minimizing_2017,claeys_floquet-engineering_2019, Hartmann_2019, passarelli_2020} that this infinite-temperature norm still provides excellent results even considering e.g. only dissipation from the ground state. Colloquially, using the Fr\"obenius norm to find the VAGP is similar to optimizing an ice cream recipe inside a very hot oven and then applying this recipe inside a freezer to efficiently prepare the ice cream. Remarkably, this procedure works amazingly well in various systems.

Rather than keeping the discussion maximally general, we will focus on a two-dimensional parameter space with controls set by $g$ and $h$ (see Eq.~\ref{TFI}), such that $\vec \lambda=(g,h)$ and analyze infinitesimal deformations $(h+\delta \cos \varphi , g+\delta \sin \varphi )$ with an infinitesimal $\delta$, such that $\unitvector = (\cos \varphi, \sin \varphi)$. The generalization of this methodology to more parameters is straightforward. Crucially, the action $S$ defined in  Eq.~\eqref{equation variational} is quadratic in the variational parameters, such that the minimization will give rise to a set of linear equations and the VAGP in an arbitrary direction will be a linear combination of the solutions corresponding to $\delta h$ ($\varphi=0$) and to $\delta g$ ($\varphi=\pi/2)$. We can write
\begin{align}
	\mathcal{A}_{\lambda}(\varphi) \equiv \unitvector \cdot \vec{\mathcal{A}}(\lambda)=  \mathcal{A}_h \cos \varphi+  \mathcal{A}_g \sin \varphi,
	\label{AGP linear}
\end{align}
in which $\mathcal{A}_h$ and $\mathcal{A}_g$ minimize the action $S_h$ and $S_g$ respectively. The Hamiltonian is set by the parameters $\vec{\lambda}$, as denoted in the subscript (where we dropped the vector notation), while the argument denotes the direction in which this Hamiltonian is varied. 
Defining 
\begin{align}
\tan 2\alpha = \frac{{\rm Tr}[\mathcal{A}_h^\dagger \mathcal{A}_g]+{\rm Tr}[\mathcal{A}_h \mathcal{A}_g^{\dagger}]}{{\rm Tr}[\mathcal{A}_h^\dagger \mathcal{A}_h]-{\rm Tr}[\mathcal{A}_g^\dagger \mathcal{A}_g]},
\end{align}
it can easily be checked that the norm of the VAGP is minimal for $\varphi=\alpha \pm \pi/2$, $\alpha \in [-\pi/4,\pi/4]$, and maximal in the orthogonal directions $\varphi =\alpha$ and $\alpha+\pi$ if $||\mathcal{A}_g|| > || \mathcal{A}_h||$, while in the other case the extrema are exchanged (see also Appendix \ref{app:derive_opt_dir}). We will call these directions optimal and orthogonal respectively. In the following sections, we will analyze the geometric structure of these directions and the resulting anisotropy as a function of $(g,h)$. Note that this also highlights that the directions set by $\varphi$ and $\varphi+\pi$ are equivalent since they correspond to the same perturbation, only with a different sign (which does not influence the norm of the VAGP).

For translationally-invariant spin-$1/2$ systems of size $L$ with periodic boundary conditions, like those described by the Hamiltonian~\eqref{TFI}, we define the $k$-body operator space $\mathcal{H}_k$, $k<L$, as the zero-momentum space of all operators having support of up to $k$ sites, where we will choose strings of Pauli matrices as basis operators: $\mathcal{H}_k  = {\rm span}(S_k)$, with
\begin{align}
	S_k=
	\{O_n| \ O_n=\sum_{p=1}^{L}   \sigma_{p}^{s_1} \sigma_{p+1}^{s_2}\cdots \sigma_{p+k-1}^{s_k}
	\},
\end{align}
where the index $n$ stands for the set $\{s_1,\dots,s_k\}$ and $\sigma^s_i$ is one of the Pauli operators $\{\sigma^x,\sigma^y,\sigma^z,1\}$ acting on the site $i$. To avoid double-counting the identity operator is excluded from the right boundary, i.e. ${s_k} \neq 1$. We will use a local variational ansatz with a fixed support:
\begin{align}
	\mathcal{A}_\lambda(\varphi)=\sum_{O_n\in S_k
} c_n(\vec{\lambda},\varphi) O_n.
	\label{variational ansatz}
\end{align}

We call this the $k$-body ansatz of the variational calculation, and solve Eq.~\eqref{equation variational} with the ansatz~\eqref{variational ansatz}. Since all operators $O_n$ are traceless and orthogonal, satisfying ${\rm Tr} (O_n O_m)/N=\mathcal D \delta_{nm}$, where $\mathcal D=2^N$ is the Hilbert space dimension, the minimization problem is straightforward and the solution is formally given by
\begin{align}
	\mathcal{A}_\lambda(\varphi)=-i\, {\rm ad}_{P_kHP_k}^{-1} \left(\unitvector  \cdot \partial_{\vec{\lambda}} {H}\right), 
	\label{solution variational}
\end{align}
where ${\rm ad}_{P_kHP_k} {\mathcal{A}} \equiv [{P}_k{H}{P}_k,\mathcal{A}]$, ${\rm ad}_{P_kHP_k}^{-1}$ is the pseudo-inverse of ${\rm ad}_{P_kHP_k}$, and ${P}_k$ is a super-operator which projects an operator onto $\mathcal{H}_k$. 

In the limit where this operator basis is complete we can consider e.g. projectors on eigenstates as basis operators, which returns the formal solution
\begin{equation}\label{eq:mat_el_A}
\mathcal{A}_{\lambda}(\varphi) = i \sum_{m\neq n}  \left| m \right\rangle \frac{ \left\langle m | \unitvector  \cdot \partial_{\vec{\lambda}} {H} |n  \right\rangle }{\epsilon_n-\epsilon_m}\left\langle n\right|,
\end{equation}
which can be checked to be equivalent to Eq.~\eqref{eq:A_Heisenberg}. 

\section{Adiabatic flow diagram of the quantum Ising model with local VAGP}
\label{sec:flow diagram}

In this section we will discuss in detail the flow diagram and the emerging physical implications for a particular, but fairly generic, quantum Ising model, which we introduced earlier in Eq.~\eqref{TFI}. We will first analyze this diagram using the VAGP obtained within the lowest-order approximation, which already yields non-trivial results. Namely, we will consider a variational manifold with support up to three sites for the VAGP. The motivation for this ansatz is that, as we discuss below, it reproduces the leading-order behavior and the most important singularities of the exact AGP near the strongest macroscopic degeneracy points. These singularities underly several key properties of the adiabatic flows and allow us to reveal the origin of special dark weakly-thermalizing states similar to those found in e.g. Ref.~\cite{wurtz_emergent_2020}. In the next section, we will then show how the results of this section are affected by adding terms with a larger support into the variational manifold. Before discussing our findings, let us mention a few properties of the Ising model that will be relevant later in the paper. 

\begin{itemize}

\item There are two integrable lines corresponding to i) $g=0$: the so-called classical Ising model with strictly local integrals of motion ($z$-magnetization for each spin) and ii) $h=0$: the transverse field Ising model, which maps to free fermions through the Jordan-Wigner transformation and which has quasi-local integrals of motion constructed from fermion bilinears \cite{calabrese_introduction_2016,essler_quench_2016}. There is an additional trivially-integrable point corresponding to $\sqrt{h^2+g^2}\to\infty$, which describes noninteracting spins. Away from these points the model is believed to be chaotic, satisfying the eigenstate thermalization hypothesis (ETH) \cite{kim_testing_2014}.

\item The ground state of the Ising model undergoes a quantum phase transition from an anti-ferromagnet corresponding to small magnetic field to a paramagnet at large magnetic field \cite{sachdev_2011}. On the integrable lines, the critical line separating the two phases terminates at the points $(h,g) = (2,0)$ and $(0,1)$. We note that changing the sign of the $ZZ$ coupling moves this phase transition line from the ground state to the most excited state. Therefore, this sign does not affect our ``infinite temperature'' flow diagram.

\item The ``classical Ising'' line $g=0$ additionally contains macroscopic (exponential) degeneracies of the spectrum at any rational value of the longitudinal field $h$. In particular, at $h=0$ and $H=ZZ$, any configuration with the same number of domain walls has the same energy, e.g. $\left| \dots\uparrow \uparrow \downarrow \dots \right\rangle$ and $\left|\dots \uparrow \downarrow \downarrow \dots \right\rangle$. At $h=2$ and $H=ZZ+2Z$, any local spin flip from a local ``down'' to ``up''  state that creates two domain walls does not change the energy of the system, e.g. $\left|\dots \downarrow \downarrow \downarrow \dots \right\rangle$ and $\left|\dots \downarrow \uparrow \downarrow \dots \right\rangle$ are degenerate. In a similar way, at other rational points of $h$ one can always find many combinations of spin flips leaving the energy of the system invariant. Finally, the $h\to\infty$ point is also macroscopically degenerate: the energy does not change under arbitrary spin flips preserving total magnetization.

\end{itemize}

\subsection{Flow diagram for the 3-body variational ansatz}

\begin{figure}[t]
	\begin{center}
           \includegraphics[width=1.\columnwidth]{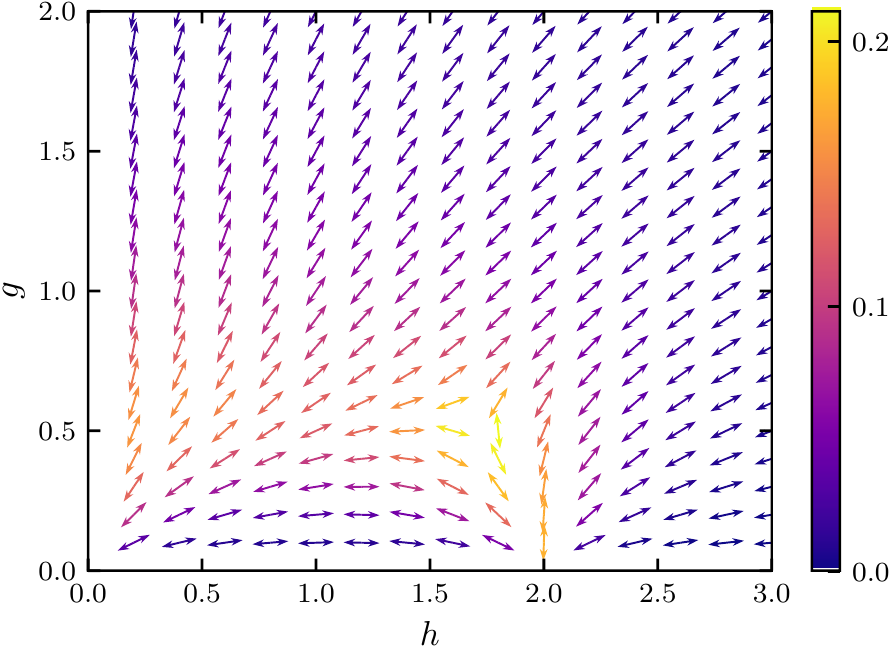}
    \caption{The flow diagram indicating the optimal direction at each point for the 3-body variational ansatz. Each point in this diagram corresponds to a Hamiltonian set by $(h,g)$ and the arrows denote the optimal direction for deformations $(\delta h, \delta g)$. Colors represent the norm of the VAGP along these optimal directions. Source flows are clearly visible at $(h,g)=(0,0)$ and $(2,0)$.}
    \label{Flow Diagram}
    \end{center}
\end{figure}

As mentioned in Section~\ref{sec:VAGP}, one can systematically define the adiabatic flow diagram by following the directions of the minimal norm of the VAGP. The resulting diagram with respect to the couplings $(h, g)$ as obtained within the 3-body variational ansatz for the VAGP is shown in the bottom half of Fig.~\ref{Flow Diagram sphere} as well as in Fig.~\ref{Flow Diagram}. Note that, on the one hand, the representation of the diagram on a sphere is more natural since all the Hamiltonians with large magnetic field are equivalent to each other up to trivial spin rotation and correspond to the same point in Fig.~\ref{Flow Diagram sphere}. On the other hand, the ``Cartesian'' representation shown in Fig.~\ref{Flow Diagram} is easier to visualize in the most interesting regime where neither $h$ nor $g$ are too large.

One can observe that the optimal flows form radial patterns centered around singularities at $(h,g)=(0,0)$ and at $(2,0)$ (as well as near $\sqrt{h^2+g^2}\to\infty$ in the spherical representation). Interestingly, these singularities lie at the endpoints of any adiabatic flow: if we start at any generic point $(h,g)$ and follow the optimal adiabatic direction, we will end up in one of these singularities. Likewise, these singular points are good starting points for quantum state preparation in e.g. quantum annealing protocols, because any point of the control space $(h,g)$ can be reached by starting at either of these singularities.  At first sight this result seems surprising: these singular points are clearly the points corresponding to large macroscopic degeneracies, where adiabatic transformations are ill-defined. Indeed, our common understanding of adiabatic transformations suggests that one should avoid situations with closing gaps between eigenstates. Thus, naively, one should generally avoid such singular points. As we will show, this reasoning only applies to the orthogonal azimuthal directions, where the norm of the VAGP becomes divergent and strong diabatic effects come into play. However, such divergences remain suppressed in the radial directions. Let us also point out that at the singular points the Hamiltonian splits into a sum of mutually commuting terms, such that its eigenstates are factorisable and thus easy to prepare.

The radial flow near $h=0$ implies that the optimal deformation of the Hamiltonian is along the instantaneous magnetic field, $(\delta h, \delta g)\propto (h,g)$. Intuitively, one can understand this result using the domain wall picture: at small magnetic fields one can think about the Ising model as a weakly interacting gas of domain walls separating regions of positive and negative magnetization. The number of the domain walls is conserved by the $ZZ$-interaction. In this manifold of states the $Z$-magnetic field plays the role of an effective linear potential and the $X$-magnetic field plays the role of the domain-wall hopping amplitude. The two terms can be combined into an effective non-interacting Hamiltonian describing these domain walls. The radial deformation of $h$ and $g$ then amounts to a simultaneous rescaling of these two parameters of the effective Hamiltonian, which does not induce diabatic transitions between the eigenstates. Similar considerations apply to the other singularity at $(2,0)$, where the effective Hamiltonian becomes the PXP model~\cite{Turner_2018} with $h-2$ playing the role of the potential and $g$ playing the role of the magnetic field. At the third degenerate point, at infinite magnetic field, the radial deformation is trivially the most adiabatic direction, since it simply amounts to rescaling the full Hamiltonian. We emphasize that, while this intuition can generally be justified by considering low-energy effective Hamiltonians, the optimal directions remain well-defined for all eigenstates. We justify this conclusion below by analytically constructing the VAGP near these points, where the radial directions are explicitly shown to be non-singular.

\subsection{State preparation along the optimal flow directions} \label{Sec better state}

\begin{figure*}
           \includegraphics[width=0.9\textwidth]{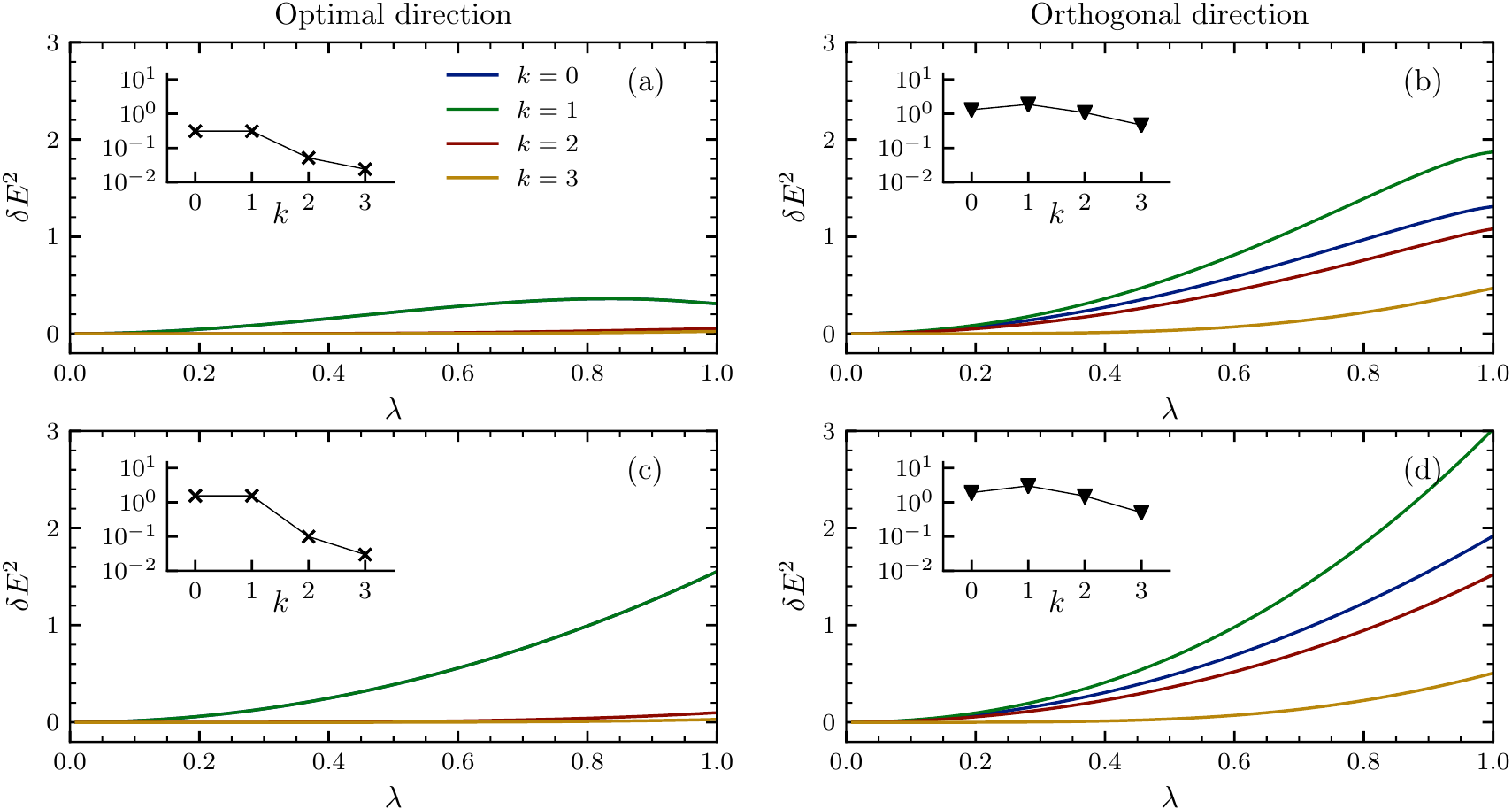}  
    \caption{Energy variance of a generic initial eigenstate evolved along the optimal direction with finite $\dot{\lambda}$ [top row (a,b)] and with infinite $\dot{\lambda}$ [bottom row (c,d)] along either the optimal direction [left column (a,c)] or the orthogonal one [right column (b,d)]. Full lines show the energy variance during the protocol as function of $\lambda(t)$, where the blue lines represent the unassisted protocol ($k=0$) and the other lines represent CD driving with $k$-body VAGPs \eqref{Ham CDD}. The inset details the energy variance at the end of the protocol as a function of $k$. The end points for both protocols are given by $(h,g) = (0.5,0.5)$, with the optimal protocol starting from $(\epsilon,\epsilon)$ and the orthogonal one from $(1-\epsilon,\epsilon)$ with $\epsilon = 10^{-2}$ (see also inset of Fig.~\ref{fig:Evardotlam}). System size is $L=12$. 
        In the unassisted protocol, the energy variance is already much smaller along the optimal directions than along the orthogonal directions. Applying local $2$-body CD driving, the energy variance drastically reduces even more along the optimal direction, while it only gradually decreases in the orthogonal direction. At infinite $\dot{\lambda}$ the state along the optimal direction can similarly be accurately approximated by a $2$-body dressing of the initial state, whereas the accuracy along the orthogonal direction only gradually increases.}
    \label{EneVar Optimal}
\end{figure*}

Before discussing the emergent features of the flow diagram in more detail, let us immediately analyze its implications for quantum state preparation. All calculations and presented diagrams hold at the operator level, so it is natural to first ask if the (operator) flows for the VAGP are representative of similar flows in the context of quantum state preparation, where only a single eigenstate is relevant. Second, if such state preparation is assisted with local counterdiabatic driving using the VAGP, a follow-up question is if the optimal directions using the VAGP are also the ones where the approximate counterdiabatic driving is maximally effective. Here, we will present numerical evidence that suggests a positive answer to these two questions.

Since we are not necessarily interested in the ground state and will consider excited states, a good measure for the proximity of any prepared state $| \psi \rangle$ to an eigenstate of the instantaneous Hamiltonian is the energy variance $\delta E$: 
\be
\delta E^2=\langle \psi | H(\vec \lambda)^2 |\psi\rangle-\langle \psi| H (\vec \lambda) |\psi\rangle^2,
\label{eq:energy_var}
\ee
where $|\psi\rangle$ is the state prepared according to a protocol following a particular, e.g. optimal, path. If the system is prepared in an exact eigenstate of $H(\vec \lambda)$, this energy variance clearly reduces to zero, whereas a non-zero value indicates how strongly this state has mixed with different-energy eigenstates. We will consider unassisted state preparation protocols and approximate CDD protocols, where the adiabatic evolution is assisted by the strictly local VAGP. In both cases we will compare different paths in control space. For the CDD protocols we numerically solve the Schr\"odinger equation using the Hamiltonian~\eqref{Ham CDD} along a given path $\vec\lambda(t)$ with $\vec{\mathcal{A}}({\vec \lambda})$ replaced by its variationally-obtained approximation. The initial state is chosen to be one of the eigenstates of the initial Hamiltonian $H$ near the middle of the spectrum, and we then compute the energy variance at the final value of $\vec \lambda$ according to Eq.~\eqref{eq:energy_var}. While the results are presented for a single (generic) eigenstate, we checked that these are representative for most eigenstates (exceptions will be discussed in Sec.~\ref{Sec dark states}).

All protocols are characterized by the total time duration $T$, where the limit of large $T$ corresponds to adiabatic evolution, while the limit of small $T$ corresponds to the instantaneous quench for the unassisted protocol and to a dressing of the initial state with the VAGP for the CDD protocol. 
We choose a smooth protocol to help eliminate diabatic effects at the protocol boundaries~\cite{kolodrubetz_geometry_2017}
\begin{align}
	\lambda(t) = \sin^2 \left( {\pi\over 2} \sin^2\left({\pi t\over 2T}\right) \right), \quad t\in[0,T] \label{CDD drive sinsin},
\end{align}
interpolating from $\lambda(0)=0$ to $\lambda(T)=1$, where we set the total protocol duration $T=2$ for concreteness, and take $(h(t),g(t)) = (h(0),g(0))+\lambda(t)(h(T),g(T))$. However, we checked that all the presented results remain qualitatively similar for other time dependences. 

\begin{figure}
\includegraphics[width=1.\columnwidth]{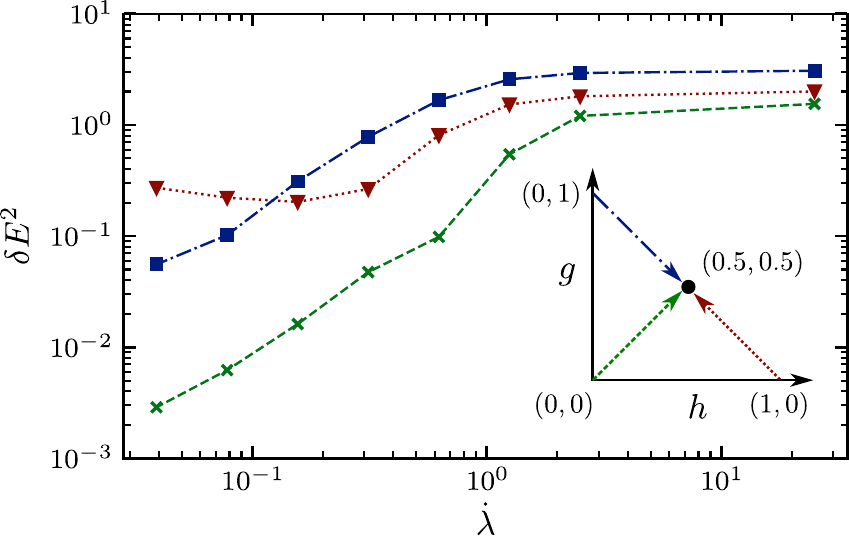}  
 \caption{Final energy variance of a generic initial eigenstate as function of protocol rate for unassisted adiabatic state preparation (see also Fig.~\ref{EneVar Optimal}). The end point is given by $(0.5,0.5)$ and initial points are given by  $(\epsilon,\epsilon)$ (optimal), $(1-\epsilon,\epsilon)$ and $(\epsilon,\epsilon-1)$ (both orthogonal), as also shown in the inset. The optimal path always outperforms the orthogonal ones.}
 \label{fig:Evardotlam}  
\end{figure}

In Fig.~\ref{EneVar Optimal} we present the resulting energy variance of the final state for different preparation protocols with the same final Hamiltonian but different initial Hamiltonians, corresponding to different directions of state preparation. For the optimal protocol, the initial point is chosen as $(h,g)=(0+\epsilon,0+\epsilon)$, with a small $\epsilon=0.01$ lifting the degeneracies of the eigenstates, which is then linearly evolved along the radial direction to the final point $(h,g)=(0.5,0.5)$ (cf. green line in the inset of Fig.~\ref{fig:Evardotlam}). This can be contrasted with the state preparation protocol along the orthogonal direction, taking as initial control parameters $(1-\epsilon, \epsilon)$ and again linearly deforming the Hamiltonian to the same final point $(0.5,0.5)$ (cf. red line). We checked that starting from another point along the orthogonal direction, namely $(\epsilon, 1-\epsilon)$, leads to similar results (cf. Fig.~\ref{fig:app:state preparation}).

In order to compare the unassisted protocols, we consider a linear ramp $\dot{\lambda} = 1/T$ and present the final energy variance for different ramp rates along different directions in Fig.~\ref{fig:Evardotlam}. It is clear that the protocol along the optimal direction generally has an energy variance that is orders of magnitude smaller than the energy variance along the orthogonal direction. Even more, when increasing $T$ (nearing adiabaticity), the energy variance for the optimal path decreases much faster, as indicated by the steeper slope in the log-log scale. Interestingly, for evolution along the sub-optimal direction starting at $(1-\epsilon, \epsilon)$, the energy variance does not decrease in the interval $0.01\leq 1/T \leq 0.1$, indicating a complicated landscape of energy level crossings. A similar situation occurs, for example, in Floquet systems \cite{weinberg_adiabatic_2017}. Still, we checked that eventually the energy variance starts decreasing again for $1/T\leq 0.005$.  

Using the calculated VAGP for approximate local CDD (see Eq.~\eqref{Ham CDD}) to improve on the unassisted protocol, panels (a) and (c) in Fig.~\ref{EneVar Optimal} show the energy variance for the CDD protocols along the optimal direction with either finite duration $T=2$ (a) or infinitely fast $T\to 0$ (c), which effectively corresponds to dressing the initial state with the VAGP. Different colors correspond to a different size of the variational ansatz for the VAGP, with the unassisted protocol included as reference. Panels (b) and (d) show related results for state preparation along the orthogonal direction. Again, it is clear from the plot that the energy variance is generally smaller for state preparation along the optimal direction. Even more, including (approximate) local counterdiabatic terms can be used to drastically reduce the energy variance along the optimal direction. We note that along the optimal direction the VAGP for 1-body ansatz is found to be exactly zero; therefore the results for $k=0$ and 1 completely overlap each other. While including the approximate counterdiabatic term along the orthogonal direction also systematically reduces the energy variance with increasing ansatz size, its effect is not as pronounced as along the optimal direction.

\subsection{Asymptotic behavior of the VAGP near singular points}    \label{Sec Anisotropy}
\begin{figure}
    	\includegraphics[width=1.0 \columnwidth]{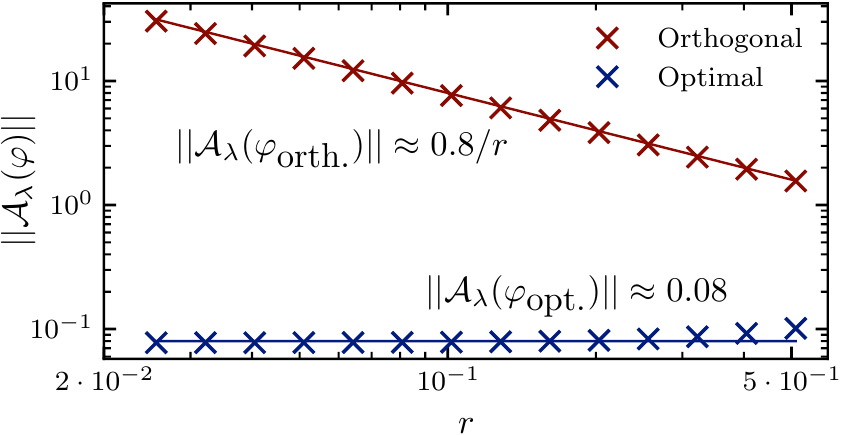}
    \caption{The norm of the VAGP for the 3-body anzatz with different $h$ and $g$. The vertical axis is the norm and the horizontal axis is $r=\sqrt{h^2+g^2}$. The ratio between $h$ and $g$ is fixed to satisfy $g=0.2\, h$. The norm in the optimal direction (blue) is nearly constant for small $r$. By contrast, the norm along the orthogonal direction (red) diverges as $O(1/r)$ as $r$ approaches to zero.} 
    	\label{Divergence2}
\end{figure}

From the structure of adiabatic flows shown in Figs.~\ref{Flow Diagram sphere} and~\ref{Flow Diagram}, it is clear that the points $(0,0)$ and $(2,0)$ play a special role, serving as sources/sinks of these flows. As already mentioned, these points also correspond to Hamiltonians with macroscopic (exponential) degeneracies in their energy spectrum. As will be discussed in this section, these points control many important properties of the AGP, including the large anisotropy between optimal and orthogonal directions and the existence of special dark/non-thermal states far from the edges of the spectrum.

In order to understand these properties, we consider perturbative expansions of the exact AGP near these two singular points. The full formalism will be developed in Sec.~\ref{sec:perturbative}, and here we will focus on the leading-order terms only. Near $(h,g)=(0,0)$, the dominant term in the perturbative expansion is given by 
\begin{align}\label{eq:agp_expansion}
\mathcal A_{\lambda}(\varphi) \approx & \frac{1}{r} {{\sin\varphi \cos\theta-\cos\varphi \sin\theta}\over 4 \cos^2\theta} (Y-ZYZ)+\dots 
\end{align}
with $r = |\vec{\lambda}|= \sqrt{g^2+h^2}$. Here the angle $\theta$ characterizes the magnetic field in the Hamiltonian $H(h,g)$ through $(h,g) = (r \cos \theta, r \sin\theta)$, whereas the angle $\varphi$ characterizes the direction in which this magnetic field is perturbed $(\delta h,\delta g) \propto  (\cos \varphi, \sin \varphi)$.

From Eq.~\eqref{eq:agp_expansion} it is clear that the AGP diverges at $(0,0)$ for a general $\varphi$. However, along the radial direction $\varphi=\theta$ the singular term exactly vanishes, indicating that the radial direction is the optimal one. It is also evident that the anisotropy between the optimal and orthogonal directions diverges near this singularity. This perturbative expansion also highlights that the variational ansatz for the VAGP minimally requires 3-body terms in order to correctly capture the singularity and the corresponding anisotropy. The increasing anisotropy as the magnetic field  goes to zero is clearly visible in the 3-body VAGP, as illustrated in Fig.~\ref{Divergence2}. In this plot we show the norm of the 3-body VAGP along the optimal and orthogonal directions as a function of $r$ at a fixed angle $\theta=\arctan(0.2)$, such that $g=0.2\ h$. The lines are the fits to the constant (optimal) and $1/r$ (orthogonal) asymptotes expected from perturbation theory. Interestingly, the perturbative scaling of the norm of VAGP extends up to a relatively large value of the coupling $r=0.4$, such that the effects from the singular point can remain important deep into the ergodic regime of the flow diagram. In Appendix \ref{app:scalingterms}, the individual weights of the terms in the expansion are compared with the scalings from perturbation theory, and it is confirmed that the dominant terms are of the form \eqref{eq:agp_expansion}.

The operator divergence can immediately be connected to the eigenstate structure of the Hamiltonian at $(0,0)$. As already noted, the energy of the model only depends on the number of domain walls, leading to macroscopic degeneracies in the eigenspectrum. The operator $Y-ZYZ$ can be seen as a `dressed' version of the spin flip operator $Y$, which however only creates a spin flip if it does not change the number of domain walls, connecting the degenerate eigenstates. This macroscopic degeneracies in $H$ and their splitting by the perturbation effectively dominate the perturbative AGP and lead to well-defined local terms.

A very similar structure emerges near the second singularity $(2,0)$, where the perturbative expansion of the exact AGP yields (see again Sec.~\ref{sec:perturbative})
\begin{equation}\label{eq:pertPYP}
\mathcal{A}_{\lambda}({\varphi}) \approx  \frac{\sin \varphi \cos \theta-\cos\varphi \sin \theta}{8r \cos^2\theta} PYP +\dots,
\end{equation}
where now $r = \sqrt{(h-2)^2+g^2}$ and $\varphi$ is again the angle characterizing the deformation $\vec {\delta \lambda}$. We introduced the notation $P$ for the projector on the down state of the spin along the $z$-direction. In the extended notation, the $PYP$ term reads
\begin{equation}
PYP=\frac{1}{4}\sum_j \left({1-\sigma_{j-1}^z}\right) {\sigma_{j}^y} \left( {1-\sigma_{j+1}^z} \right).
\end{equation}
Same as near the $(0,0)$-singularity, the AGP diverges as $r\to 0$ except in the radial direction $\phi=\theta$. Therefore the AGP again becomes infinitely anisotropic in the limit $r\to 0$. This singularity is precisely reflected in the flow diagram indicating that the optimal directions are radial.

Interestingly, and not accidentally, the leading-order singularity of the AGP is nothing but the generator of spin rotations of the effective low-energy $PXP$ model emerging near the $(2,0)$ point~\cite{Turner_2018}. This model was already shown to satisfy highly unusual properties, including the existence of weakly thermalizing quantum scar states~\cite{Turner_2018} and the existence of nearby integrable deformations of the Hamiltonian~\cite{Khemani_2019}. In the next section, we will show that some (and probably all) unusual properties of this model are encoded in the exact AGP and can be observed in its local variational approximation.

Since it was recently noted that the AGP generates the effective Schrieffer-Wolff Hamiltonian, it is also worthwhile to note that the effective PXP Hamiltonian can be obtained by performing the Schrieffer-Wolff transformation using the VAGP \cite{wurtz_variational_2020}.

\subsection{Many-body dark states} \label{Sec dark states}
\begin{figure*}
	\includegraphics[width=0.8\textwidth]{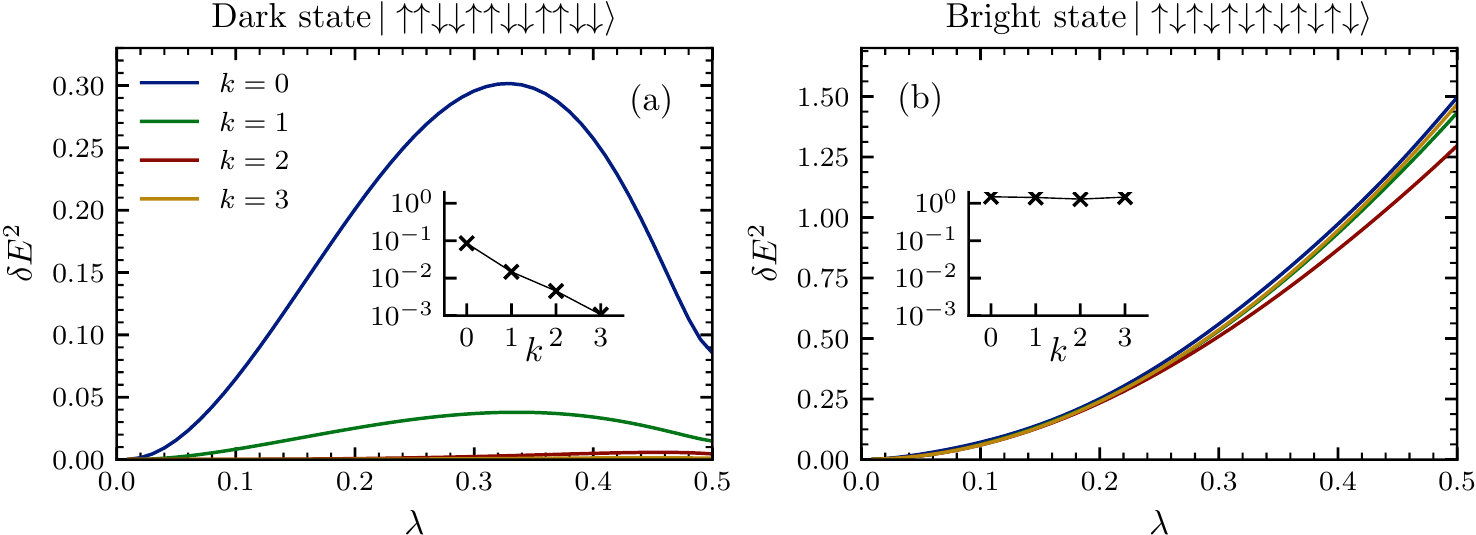}
    \caption{Energy variance of a dark (a) and a bright (b) eigenstate as function of $\lambda$. Blue lines represent the unassisted protocol ($k=0$) and the other lines represent CD driving with a $k$-body VAGP using a sin-square ramp \eqref{CDD drive sinsin}. Inset details the final energy variance as function of $k$. Note the different vertical scales in both figures. The starting point of the protocol is $(h,g) = (2,0)$ and the final point is $(h,g)=(2,0.5)$. System size is $L=12$. Initial states are the ``dark state" $\state{\psi_1}=\state{\uparrow \uparrow \downarrow \downarrow \uparrow \uparrow \downarrow \downarrow \uparrow \uparrow \downarrow \downarrow }$ (a) and the N\'eel state  $\state{\psi_2}=\state{\uparrow \downarrow\uparrow \downarrow \uparrow \downarrow \uparrow \downarrow \uparrow \downarrow \uparrow \downarrow  }$ (b). Even for the unassisted protocol, the final  energy variance is much smaller for $\state{\psi_1}$ than for $\state{\psi_2}$. Introducing a local counterdiabatic term rapidly decreases the final energy variance in the dark states, whereas the energy variance remains largely unchanged in the N\'eel state (see insets).} 
    \label{Dark states}
\end{figure*}

The local structure of singularities of the AGP near the macroscopically degenerate points also allows for the existence of special states that are simultaneously eigenstates of the Hamiltonian and are annihilated by (or are possibly other eigenstates of) 
the leading divergent part of the AGP. From Eq.~\eqref{eq:mfHam} it is clear that such states should be largely immune to {\em any} time-dependent protocols $\vec \lambda(t)$. They are thus approximately {\em dark states}. 

Let us start by analyzing such states near the singularity at $(2,0)$. From Eq.~\eqref{eq:agp_expansion} it follows that the divergent part of the AGP in any direction except the radial one scales as
\[
\mathcal{A}_{\rm s}\propto {1\over r}PYP.
\]
We can readily see that $\mathcal{A}_{\rm s}$ has many zero eigenstates that are simultaneous eigenstates of $H$ at $r=0$. An example of such a state is 
\begin{align}
 	\state{\psi_1}=\state{\uparrow \uparrow \downarrow \downarrow \uparrow \uparrow \downarrow \downarrow \uparrow \uparrow \downarrow \downarrow }.
\end{align}
There are (exponentially) many other such dark states, which can e.g. be obtained by increasing the length of the domains of $|\uparrow\rangle$ spins. From Eqs.~\eqref{eq:mfHam} and \eqref{comoving}, the time evolution of such a $\state{\psi_1}$ in the co-moving basis under an arbitrary time-dependent protocol is given by
\begin{align}
\label{eq:schr_eq_dark_states}
 	i{\partial \over \partial t}\state{\psi_1}
 	&=(H-\dot{\lambda} \mathcal A_{\rm n})\state{\psi_1},
\end{align}
where $\mathcal A_n$ is the remaining non-divergent part of the AGP as
\begin{align}
\mathcal{A}_{\lambda} \state{\psi_1} = \left(\mathcal{A}_{\rm s}+\mathcal{A}_{\rm n}\right)\state{\psi_1} = \mathcal{A}_{\rm n}\state{\psi_1}.
\end{align}
We see that the state $|\psi_1\rangle$ is unaffected by the term $\mathcal{A}_{\rm s}$, the main source of diabatic excitations in general states, and is thus only weakly excited. Because this statement is general and is not tuned to the details of the protocol $\vec\lambda(t)$, this state approximately behaves as a many-body dark state. The remaining non-divergent terms $\mathcal{A}_{\rm n}$ entering Eq.~\eqref{eq:schr_eq_dark_states} can be further suppressed by means of local CDD. As we show in Sec.~\ref{sec:perturbative}, $\mathcal A_{\rm n}$ has a well defined expansion in terms of local operators and thus the dark states only acquire local dressing near singularities and remain highly nonthermal (with e.g. low entanglement entropy) even far from the singularity, in the ergodic regime.

To demonstrate the advantage of the many-body dark state in the context of quantum state preparation, we consider a CD protocol with the VAGP, starting at the singular point $(2,0)$ and subsequently increasing the transverse magnetic field up to the point $(2,0.5)$. We consider two scenarios, starting with two different initial states, a dark state $\state{\psi_1}$ and a bright (non-dark) N\'eel state that is not annihilated by the singular part of the AGP:
\begin{align}
\state{\psi_2}=\state{\uparrow \downarrow\uparrow \downarrow \uparrow \downarrow \uparrow \downarrow \uparrow \downarrow \uparrow \downarrow  }	.
\end{align} 
We choose the protocol given by Eq.~\eqref{CDD drive sinsin} with protocol duration $T=1$. 
The results of the simulations are shown in Fig.~\ref{Dark states}.
In Appendix \ref{app:darkstates}, we analyze the dressing of less symmetric dark and bright initial states, and show that they exhibit a very similar qualitative behavior. 

Even for the unassisted protocol (blue lines), we can already see in the figure that the energy variance of the dressed dark state is a factor of $20$ smaller than that of the bright state. This ratio quickly increases if we increase the protocol duration. The difference between the dark and bright states becomes even more pronounced in the presence of the local CD term. We see that the energy variance of the bright N\'eel state $\state{\psi_2}$ is almost unaffected by the counterdiabatic term, only decreasing from $1.496$ to $1.468$ as we go from the unassisted protocol to the CDD with the $3$-body ansatz. On the other hand, the energy variance of the prepared dark state reduces from $0.085$ (unassisted) in the unassisted protocol to $0.001$ ($3$-body CDD) for the dark state. Such a small energy variance implies that the prepared state is very close to an eigenstate of the system. The fact that this state is prepared in a short time $T=1$ using a local CD Hamiltonian also implies that this state is nonthermal, e.g. it exhibits area law entanglement.

It is easy to check that the dark states, i.e. the zero-energy eigenstates of the $PYP$ Hamiltonian, are simultaneously the zero-energy eigenstates of the low-energy effective $PXP$ Hamiltonian. Interestingly, the AGP allows us to find these special states without prior knowledge of the effective Hamiltonian.

One can similarly analyze the structure of the AGP near the other singularity at $(0,0)$. From Eq.~\eqref{eq:agp_expansion} it follows that the divergent part of the AGP is given by
\[
\mathcal{A}_{\rm s}
\propto Y-ZYZ.
\]
This operator clearly annihilates two pairs of states: i) fully-polarized states $\state{\uparrow \uparrow\dots \uparrow \uparrow}$ and $\state{\downarrow \downarrow\dots\downarrow\downarrow}$ and ii) the two N\'eel states $\state{\uparrow \downarrow\uparrow \downarrow\dots \uparrow \downarrow}$  and $\state{\downarrow\uparrow \downarrow\uparrow\dots \downarrow \uparrow}$. The two N\'eel states are clearly the degenerate ground states, such that it is not surprising that they can be efficiently dressed locally as we introduce a nonzero finite magnetic field. The two ferromagnetic states are the most excited states, i.e. the states with maximal energy. As we turn on the $Z$-magnetic field, one of the polarized states remains the most excited state -- it is again not surprising that this state can be locally dressed. However, the second polarized state quickly enters the energy continuum and yet, because it is annihilated by $\mathcal A_s$, it only weakly hybridizes with other states and remains highly non-thermal. This dark state was recently discovered in Ref.~\cite{wurtz_emergent_2020} (cf. Fig. 4 there) as a state with anomalously low entanglement. Interestingly, in this case the ground and most excited states can be immediately determined as the zero states of the AGP, without any need to diagonalize the full Hamiltonian.

\section{Flow diagram with the higher order variational ansatz}\label{sec:bigger ansatz}
\begin{figure}
	\begin{center}
           \includegraphics[width=\columnwidth]{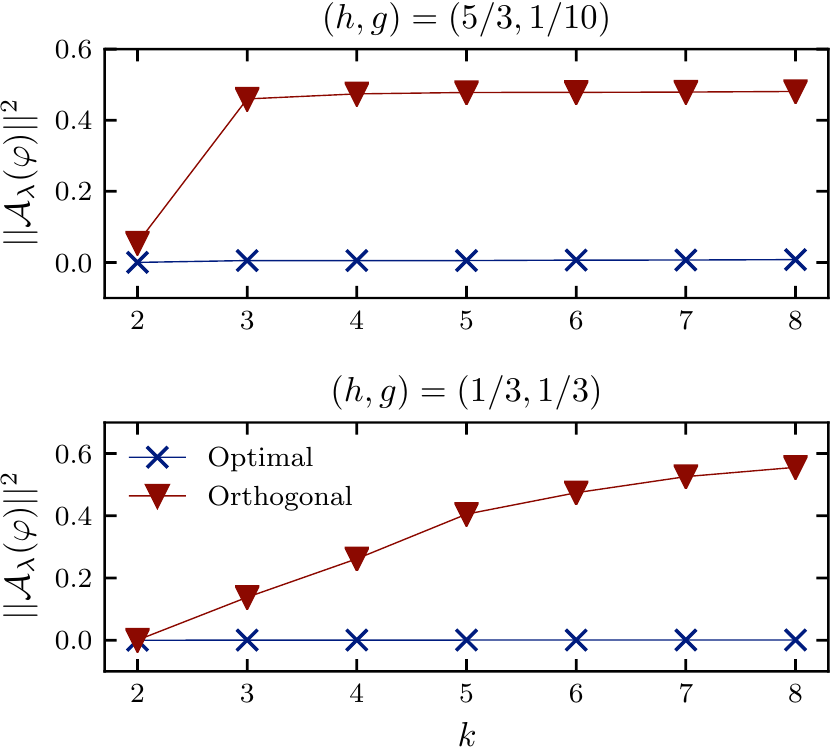}
    \caption{The norm of the VAGP for different ansatz sizes $k$.  The upper figure corresponds to $(h,g)=(5/3,1/10)$ and the lower figure to $(1/3, 1/3)$. The blue (red) points are the norms of the AGP in the optimal (orthogonal) direction. The AGP shows a high degree of anisotropy for $k\gtrsim 3$.}
    \label{Optimal Directions}
    \end{center}
\end{figure}

\subsection{Scaling of the VAGP norm with the ansatz size}

Having analyzed the emerging adiabatic flow diagram within the 3-body variational ansatz, we now consider what happens when we increase the support of the VAGP to more than three sites. First, we study how the norm of the VAGP changes with the increasing ansatz size $k$. A slow increase of $||\mathcal A_\lambda||$ with $k$ indicates that increasing the support of the ansatz only has a small effect on the VAGP, such that its local approximation is stable and accurate. Conversely, a fast increase of $||\mathcal A_\lambda ||$ with ansatz size would indicate that the exact AGP is highly non-local and the local variational ansatz is not very stable. In Fig.~\ref{Optimal Directions}, we analyze the norms of the VAGP in the optimal (blue) and orthogonal (red) directions at two different sets of couplings: $(5/3,1/10)$ (top) and $(1/3, 1/3)$ (bottom). The first point is close to the $g=0$ classical Ising line and relatively far from the singular points, whose structure is explained below. The second point is dominated by its proximity to the $(0,0)$ singularity, but it is not too close to it. In both cases we observe a large anisotropy between the optimal and orthogonal directions starting from $k=3$. In particular, we see that the norm of the VAGP in the orthogonal direction rapidly increases to a large value as $k$ reaches $3$ and then remains relatively flat for the first set of couplings, and increases more gradually with $k$ for the second set of couplings. In both cases the AGP norm in the optimal direction increases slowly with $k$. As we will show below, when we keep increasing the ansatz size, new singularities affecting the VAGP start to emerge. These singularities can discontinuously change the optimal direction, at the same time drastically reducing the anisotropy of the AGP.

\subsection{Emergence of new singular points}

\begin{figure}
	\begin{center}
	\includegraphics[width=1.0\columnwidth]{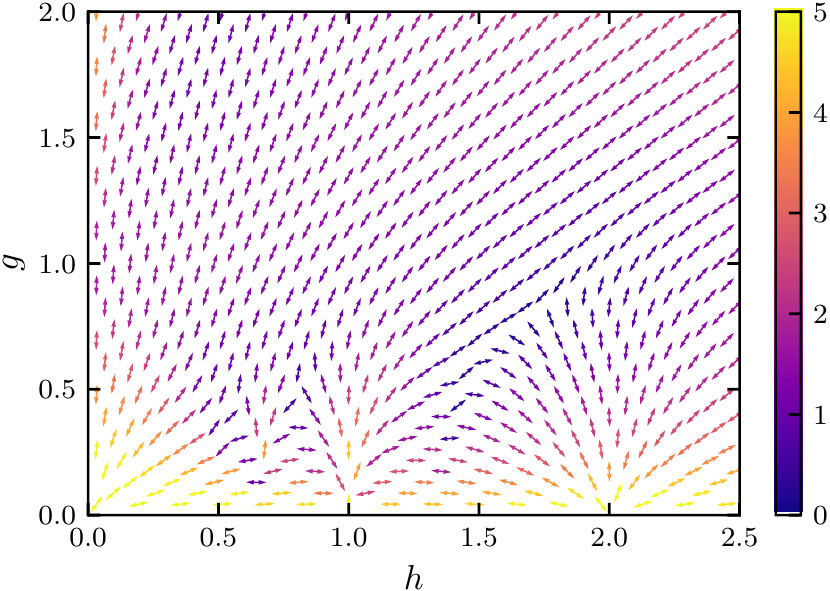}
    \caption{The flow diagram indicating the optimal direction at each point for the 8-body ansatz. Each point in this diagram corresponds to a Hamiltonian set by $(h,g)$ and the arrows denote the optimal direction for deformations $(\delta h, \delta g)$. The color now represents the logarithm of the ratio of the norm in the optimal direction over that in the orthogonal direction, ranging from blue (nearly anisotropic) to yellow (highly isotropic). Source flows are clearly visible not just at $(h,g)=(0,0)$ and $(2,0)$, but also at $(1,0)$ and $(2/3,0)$. }
    \label{Flow diagram 7body}
    \end{center}
\end{figure}

In Sec.~\ref{sec:perturbative} and Appendix~\ref{app:perturbative}, where we discuss the perturbative expansion of the AGP for small values of $g$, we show that new singularities emerge in correspondence with the degenerate points along the line $g=0$ when increasing the support of the VAGP ansatz. For example, in the second-order approximation a new singularity at $h=1$ appears, in the third-order a singularity appears at $h=2/3$, etc. These singularities correspond to correlated rearrangements of spins leaving the energy of the unperturbed Hamiltonian invariant, which correspondingly involve longer and longer strings of operators in the AGP. In other words, distinguishing degenerate states from each other through local operators requires operators with increasing support, which will arise at higher orders in the perturbative expansion. For example, the leading-order singular term near $(h,g)=(1,0)$ reads (see Appendix~\ref{app:perturbative}):
\begin{align}\label{eq:Adiv_h1}
\mathcal{A}_\lambda(\varphi) =& \frac{\sin \theta}{32 \cos^2\theta} \left(\sin \theta \cos \varphi - 2 \cos \theta \sin \varphi\right)\nonumber\\
&\quad \times  P(XY+YX)P+\dots,
\end{align}
where we now parametrize the magnetic field in the Hamiltonian as $(h,g) = (1+r\cos\theta, r \sin \theta)$ and the direction in which we perturb is again given by $(\delta h, \delta g) \propto (\cos\varphi, \sin \varphi)$. One can readily see that this singularity is not radial and only develops around $\theta=\pi/2$. It is weaker than the previously-analyzed singularities at $(0,0)$ and $(2,0)$ due to the absence of the $1/r$ divergent prefactor (cf. Eqs.~\eqref{eq:agp_expansion} and~\eqref{eq:pertPYP}), so the divergence is confined to a narrow angular region. The optimal direction near $\theta=\pi/2$ is again the one where the divergent part of $\mathcal{A}_\varphi$ vanishes, corresponding to $2\cot\varphi = 2\cot \theta$, which implies that  $\delta\varphi\approx 2\delta \theta$, where $\delta \varphi=\pi/2-\varphi,\; \delta\theta=\pi/2-\theta$. Hence, the optimal direction is no longer radial, except exactly at the singularity, where $\theta=\pi/2$. Since the operator part of the diverging contribution to the AGP contains four-body operators, this singularity will only manifest in the VAGP if we use a $4$-body ansatz or higher. This is exactly what is shown in Fig.~\ref{Flow Diagram sphere}, where the flow diagram for the 5-body ansatz contains sources/sinks at both the 3-body singularities $(0,0)$ and $(2,0)$ and the additional singularity $(1,0)$.

\begin{table}[t]
	\begin{tabular}{|c |c | c|} \hline
		$h$  & Operator & Degenerate states\\ \hline
		0 & $Y-ZYZ$ & $\state{\cdots\up \up \dw\cdots} \leftrightarrow \state{\cdots\up \dw \dw \cdots}$  \\ \hline 
		2 &	$PYP$ & $\state{\cdots\dw \up \dw\cdots} \leftrightarrow \state{\cdots\dw \dw \dw\cdots}$  \\ \hline 
		1 & $P(XY+YX)P$ & $\state{\cdots\dw \up \up \dw\cdots} \leftrightarrow \state{\cdots\dw \dw \dw \dw\cdots}$  \\ \hline 			
		\raisebox{-0mm}{$2\over3$}&
		\raisebox{-0mm}{$P(YXX+XYX$} & 
		\raisebox{-0mm}{$\state{\cdots\dw \up \up \up \dw\cdots} \leftrightarrow \state{\cdots\dw \dw \dw \dw \dw\cdots}$}  \\ 
		&\raisebox{0mm}{$\ \ \ \ \ \ +XXY-YYY)P$} &\\
		\hline 
		$\vdots$ &$\vdots$ &$\vdots$ \\ \hline
	\end{tabular}
	\caption{Singular contribution to the VAGP at different singular points $(h,0)$ and corresponding spin flips conserving the energy. Operators with increasing support lead to weaker divergences appearing in higher-order perturbative terms at rational values of $h$.}
	\label{table divergent terms}
\end{table}

Increasing the support of the ansatz will lead to additional singularities, which can be captured in higher-order terms in the perturbative expansion. As such, higher-order singularities will become even more suppressed in orders of $r$, such that they will manifest themselves only some distance away from the degenerate $g=0$ line. 
In Fig.~\ref{Flow diagram 7body}, we show the flow diagram for the 8-body variational ansatz. The arrows again indicate the optimal directions, and the color now represents the anisotropy, i.e. the ratio of the VAGP norm along the optimal and the orthogonal directions, with yellow indicating a higher anisotropy. New singularities at $h=1$ and $h=2/3$ become visible in this plot, accompanied by additional, non-radial, structures around them. 

Clearly, the leading-order singular term can be singled out either perturbatively or variationally. As argued above, the corresponding operators should connect states that are exactly degenerate at the corresponding singular point. In Table~\ref{table divergent terms} we summarize these leading-order operators and illustrate how they connect degenerate states through correlated spin flips, inducing both the macroscopic degeneracies in the eigenspectrum and the divergences in the VAGP.

We note an interesting feature following from Fig.~\ref{Flow diagram 7body}: as we increase the size of the variational ansatz, in some regions the optimal direction can switch. This is most clearly visible near the point $h=1$ and small $g$. Within the 3-body ansatz, the optimal direction is nearly horizontal (cf. Fig.~\ref{Flow Diagram}), while in the higher-body ansatz ($k>4$) the optimal direction is nearly vertical (cf. Fig.~\ref{Flow diagram 7body}). This discontinuity indicates that it is impossible to improve the accuracy of the VAGP in the horizontal direction by increasing the support of the variational ansatz: the new singularity prevents us from doing so. The only way to continue improving local state preparation is to change the direction. It is clear that such a sudden change should introduce some ambiguity in finding the optimal path in the space of couplings in the vicinity of the singularity. Indeed, we see that regions of small anisotropy surround the singularity at $(1,0)$ -- in such regions the difference between the optimal and the orthogonal directions is less pronounced.

\section{VAGP and approximately conserved operators}\label{sec: conserved operator}

As we discussed above, the VAGP for deformations along the direction $\lambda_j$ is found by minimizing the norm of the  operator $G_j$ (cf. Eqs.~\eqref{eq:G_def} and~\eqref{equation variational}). If the VAGP is exact, then $G_j$ is a conserved operator conjugate to the direction $\lambda_j$. However, for an approximate VAGP, $G_j$ is only approximately conserved because it has a non-zero commutator with the Hamiltonian. It is clear that the norm of the commutator $[G_j,H]$ is a measure for the accuracy of this approximate conservation law: the smaller the norm, the better the conservation law. In some sense, this norm serves as a proxy to the magnitude of the difference between the exact and the local variational AGP. If this difference is small, we can simultaneously implement accurate local counterdiabatic driving and construct a local nearly-conserved operator. These qualitative considerations are indeed correct, as we show below by analyzing the accuracy of such conservation laws in the optimal directions at different couplings and different ansatz sizes.

\begin{figure*}[ht]
	\begin{center}
           \includegraphics[width=0.8\textwidth]{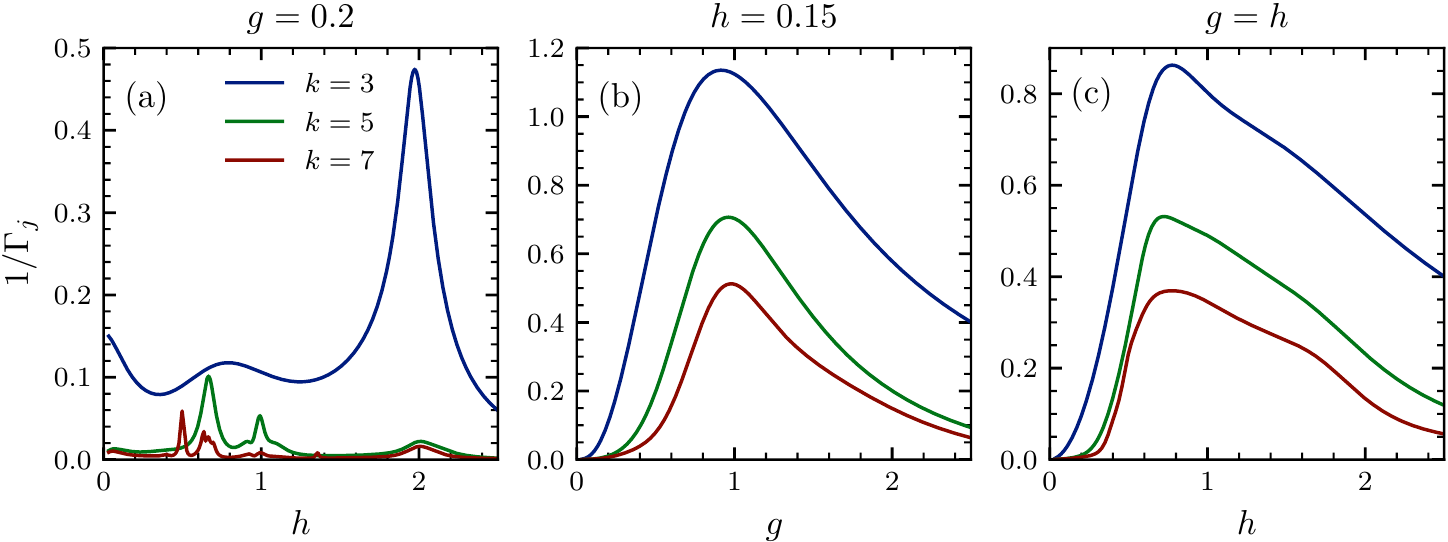}
    \caption{Inverse lifetime $\Gamma_j$ for nearly-conserved operators constructed using the VAGP along the optimal directions for $k=3,5,7$. Increasing the support of the VAGP increases the lifetime. The maxima are observed near the quantum critical point $(0,1)$ and singular point $(2,0)$, which indicates that the infinite temperature VAGP is still ``aware" of the zero temperature quantum critical behavior. Inverse lifetime along (a) a line with fixed $g$, (b) a line with fixed $h$, and (c) the line $g=h$.} 
    \label{[HG]}
    \end{center}
\end{figure*}

A more convenient and physical measure characterizing the accuracy of the conservation law  is the lifetime of $G_j$ measured in an eigenstate $\left|n\right\rangle$ of the Hamiltonian. The latter can be computed from the short-time expansion of the connected non-equal time correlation function~\cite{kim_slowest_2015}:
\begin{align}
&{1\over 2} \langle n | G_j(t) G_j(0) + G_j(0) G_j (t) |n\rangle_c \nonumber\\
&= \langle n| G_j^2(0) |n\rangle_c-{t^2\over 2} |\langle n | [H, G_j]^2 |n\rangle_c|+\mathcal O(t^4).
\end{align}
From this expansion one can define a state-averaged normalized decay rate (inverse lifetime) for the operator $G_j$ as
\begin{align}
	\Gamma_j^2={\left|{\rm Tr}\left[[H,G_j]^2\right]\right| \over {\rm Tr}[G_j^2]} = \frac{||[H,G_j]||^2}{||G_j||^2}. 
	\label{eq [HG]}
\end{align}
A small decay rate indicates that the operator $G_j$ is nearly conserved, at least up to times of the order $1/\Gamma_j$.  For the exact AGP, obviously, $\Gamma_j=0$.

In Fig.~\ref{[HG]} we show the lifetimes of the operators $G_j$ computed in the optimal direction, i.e. the direction shown by arrows in Fig.~\ref{Flow diagram 7body}, as a function of i) $h$ at fixed $g=0.2$ (panel a);  ii) as a function of $g$ at fixed $h=0.15$, and iii) as a function of the total magnetic field along the diagonal direction $h=g$. Different lines on each panel refer to different ansatz sizes. In all the cases we chose the direction $\lambda_j$ to be the optimal one for the corresponding ansatz size. In the panel (a), showing $\Gamma_j$ as a function of $h$ at a fixed small value of $g$, we see several characteristic features. First of all, it is clear that increasing the ansatz size increases the lifetime of the nearly-conserved operators. Furthermore, the decay rate exhibits non-monotonic peaks at the singular points of the AGP.  As we increase the ansatz size $\Gamma_j$ becomes more sensitive to the higher-order singularities. Thus the effect of the singularity near $h=2$ is very strong at $k=3$, i.e.~at the 3-body ansatz level but becomes very small for larger $k$. This picture is consistent with our previous analysis, suggesting that the divergent contributions to the VAGP, corresponding to leading singularities, are local and as such can be eliminated by the local VAGP. Higher-order singularities then require a VAGP with increasing support. Another very interesting feature emerges if we analyze the dependence of $\Gamma_j$ on $g$ at fixed small $h=0.15$ [panel (b)]. Namely, the decay rate exhibits a clear maximum near $g=1$, corresponding to the quantum critical point at zero temperature~\cite{sachdev_2011}. Interestingly, the maximum in $\Gamma$ is clearly pronounced despite the fact that we analyze the operator lifetimes at infinite temperature, where static observables do not exhibit any signatures associated with  criticality, consistent with recent results from Ref.~\cite{wurtz_emergent_2020}. At $h=0$, i.e.~in the limit of the integrable transverse field Ising model, this result is known from prior work~\cite{del_campo_assisted_2012, kolodrubetz_geometry_2017}. The plot shown in Fig.~\ref{[HG]} suggests that, even if the integrability is broken, the maximum of $\Gamma_j$ remains well-defined and again highlights how temperature plays a much less important role when we define quantum criticality through the diabatic response encoded in the AGP.

\section{Perturbative expansion}
\label{sec:perturbative}

In this final section we present a derivation of the divergences appearing in the VAGP by developing a perturbative expansion of the exact AGP  in small $g$ near $g=0$, i.e. near the classical Ising limit, using the integral representation of the AGP given by Eq.~\eqref{eq:A_Heisenberg}. We will outline only a sketch of the derivation here and provide some key results, further details of all calculations can be found in Appendix~\ref{app:perturbative}.

We will denote the Hamiltonian at the solvable point $g=0$ as $H_0 = ZZ + h Z$ and find a perturbative expansion for $\mathcal{A}_{\lambda}$ at $H=H_0+gX$ in powers of $g$ for general $\partial_{\lambda} H$,
\begin{align}
\mathcal{A}_{\lambda} = \mathcal{A}_{\lambda}^{(0)}+g \mathcal{A}_{\lambda}^{(1)}+ \mathcal{O}(g^2).
\end{align}
The first-order contribution can be found by setting $g=0$ in Eq.~\eqref{eq:A_Heisenberg},
\begin{align}
\label{eq:A_Interaction_0}
\mathcal A_\lambda^{(0)} = -{1\over 2}\lim_{\epsilon\to 0^+} \int_{-\infty}^{\infty} dt\, {\rm sgn}(t)\, \mathrm e^{-\epsilon |t|} (\partial_\lambda H^{(0)})(t),
\end{align}
where any time-dependence is taken to be in the interaction picture, $(\partial_\lambda H^{(0)})(t) \equiv e^{iH_0t}(\partial_{\lambda}H) e^{-iH_0t}$. The next order can be found by taking the derivative w.r.t.~$g$ in Eq.~\eqref{eq:A_Heisenberg}, 
\begin{align}
\label{eq:A_Interaction_1}
\mathcal A_\lambda^{(1)} = -\frac{i}{2}\lim_{\epsilon\to 0^+} \int_{-\infty}^{\infty} dt\, {\rm sgn}(t)\, \mathrm e^{-\epsilon |t|}  \left[\chi(t),(\partial_{\lambda}H^{(0)})(t)\right],
\end{align}
where
\begin{equation}
\chi(t) = \int_{0}^t d\tau\, X(\tau), \quad X(t) = e^{iH_0t}X e^{-iH_0t}.
\label{eq:X_t_def}
\end{equation}
 In order to simplify the notations we use $X(t)$ instead of $X^{(0)}(t)$. Higher-order terms can be found by taking higher-order derivatives of Eq.~\eqref{eq:A_Heisenberg}, leading to an iterative evaluation scheme. We will only analyze the first two orders here.

We will separately calculate the dominant terms for $\partial_{\lambda}H = X$ and $\partial_{\lambda}H=Z$, yielding  $\mathcal{A}_g$ and $\mathcal{A}_h$ correspondingly. Given a general perturbation $(\delta h, \delta g) \propto (\cos\varphi, \sin \varphi)$, we can write $\mathcal{A}_{\lambda}(\varphi) = \cos\varphi \mathcal{A}_h + \sin \varphi \mathcal{A}_g$. 

Given that $H_0=ZZ+hZ$, in the interaction picture $Z^{(0)}(t) = Z$ is time-independent, and hence $\mathcal{A}_h^{(0)}=0$. For $\mathcal{A}_g$, we need to first evaluate $X(t)$, which can be done analytically (see Eq.~\eqref{eq:X(t)_Ising} and Appendix~\ref{app:perturbative}). It represents a sum of eight different independent operators with support up to $k=3$ with time-dependent coefficients. For $h\neq 0,2$ the integral of $X(t)$ is well behaved in the limit $\epsilon\to 0$ and we can find
\begin{align}
\mathcal A_g^{(0)}=&{1\over 2h}{2-h^2\over 4-h^2} Y+{1\over 2 (4-h^2)} (YZ+ZY) \nonumber\\
&\qquad-{1\over h (4-h^2)} ZYZ.
\label{eq:A_g_0}
\end{align}
This expression clearly diverges at $h=0$ and $h=2$. Collecting the diverging terms near these singularities, we recover the expressions quoted earlier (Eqs.~\eqref{eq:agp_expansion} and~\eqref{eq:pertPYP}) in the limit $\varphi\to \pi/2$ and $\theta\to 0$.

Exactly at the singular points the divergent terms commutes with the Hamiltonian $H_0$ and can be subtracted from the AGP. This sudden discontinuity is not accidental, since the direction along $g$ becomes exactly radial at the singular points, which is optimal. The cancellation of divergences also follows from Eq.~\eqref{eq:X(t)_Ising} and arises from the fact that the limits $\epsilon\to 0$ and $h\to 0,2$ do not commute. An explicit evaluation of Eq.~\eqref{eq:A_Interaction_0} at $h=0$ returns:  
\begin{align}
\mathcal A_g^{(0)}={1\over 8}(YZ+ZY).
\end{align}
similarly at $h=2$ we find
\begin{align}
\mathcal A_g^{(0)}={5\over 32} Y+{1\over 32} (YZ+ZY)-{3\over 32} ZYZ.
\end{align}

The first non-vanishing contribution to $\mathcal{A}_{h}$ is $\mathcal{A}_h^{(1)}$, which can be immediately obtained from Eq.~\eqref{eq:A_Interaction_1} (see again Appendix~\ref{app:perturbative} for details):
\begin{eqnarray}
\mathcal{A}_h^{(1)} = -\frac{1}{(h^2-4)^{2}}\Bigg(\frac{h^4-2h^2+8}{2h^2}Y +\frac{3h^2-4}{h^2} ZYZ\nonumber\\
- h (ZY+YZ)\Bigg). \quad 
\label{eq:A_h_1}
\end{eqnarray}
In a similar fashion, one can compute an exact analytic expression for $\mathcal{A}_g^{(1)}$, showing the emergence of the new singularity at $h=1$. This expression is rather long, so is is only explicitly given in the Appendix~\ref{app:perturbative}. 

\begin{figure}[ht]
	\begin{center}
           \includegraphics[width=1.\columnwidth]{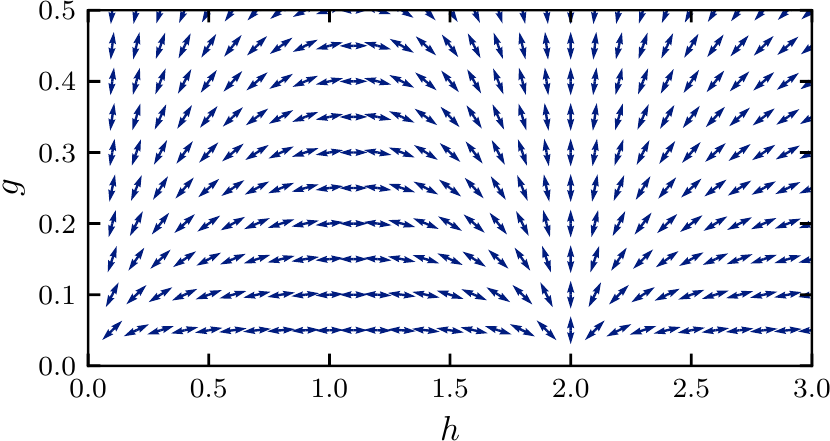}
    \caption{Flow diagram of the first-order perturbative calculation as given by Eq.~\eqref{phi analytic}. Two sources/sinks of the flows are observed at $(h,g)=(0,0)$ and $(2,0)$, reproducing the results shown in Fig.~\ref{Flow Diagram}.} 
    \label{fig flow analytic}
    \end{center}
\end{figure}

Interestingly, while formally $\mathcal A_h^{(1)}$ is obtained as a higher-order term than $\mathcal A_g^{(0)}$, it contains the same type of singularities at $h=0$ and $h=2$. Moreover, it also only contains terms with support of up to three sites: both these terms will appear in, e.g. the $3$-body variational ansatz. Physically, $\mathcal A_h^{(1)}$ plays the same role as $\mathcal A_g^{(0)}$ because both appear as the leading non-vanishing contributions to the AGP in the perturbative expansion. For this reason it suffices to analyze the following ``leading order'' perturbative AGP:
\be
\mathcal A_{\lambda}(\varphi) \approx \cos\varphi\, \mathcal A_h^{(1)} +\sin\varphi\, \mathcal A_g^{(0)}.
\label{eq:A_gh_0}
\ee
As we will show next, the AGP in this form allows us to understand key features of the adiabatic flows near the singularities at $(0,0)$ and $(2,0)$. Using Eqs.~\eqref{eq:A_g_0} and~\eqref{eq:A_h_1}, we can minimize the norm of the perturbative AGP~\eqref{eq:A_gh_0} with respect to $\varphi$ and find the optimal direction as
\begin{align}
\tan(2\varphi)
=\frac{2 g \left(h^6+24 h^2-32\right)}{h (h^2-4) \left(h^4-2 h^2+8\right)} + O(g^2).
\label{phi analytic}
\end{align}
The corresponding perturbative flow diagram is shown in Fig.~\ref{fig flow analytic}. It is clearly highly similar to the variational flow diagram obtained for the $3$-body variational ansatz (cf. Fig.~\ref{Flow Diagram}), confirming how the (numerically straightforward) variational approach is able to identify the most important local contributions to the AGP. It is easy to check that from Eq.~\eqref{eq:A_gh_0} we can recover the asymptotic behavior of the AGP close to the singularities at $(0,0)$ and $(2,0)$ (cf.~Eqs.~\eqref{eq:agp_expansion} and~\eqref{eq:pertPYP}). 

\section{Conclusions}\label{sec:conclusion}

We developed a general approach for analyzing the adiabatic landscape in systems described by a family of Hamiltonians characterized by several controls (couplings). This approach is based on minimizing the norm of the local variational adiabatic gauge potential, which serves as the local generator of adiabatic transformations. 

We applied this method to a one-dimensional Ising model in the presence of both transverse and longitudinal fields. In this model we determined the optimal directions as those where the norm of the adiabatic gauge potential is minimal, which can be used to immediately define continuous paths along which diabatic effects are suppressed (cf.~Fig.~\ref{Flow Diagram sphere}). Along these optimal paths one can design highly efficient local and experimentally feasible counterdiabatic driving protocols. These paths are also useful for various other applications, including finding local nearly-conserved operators, dressed elementary excitations such as quasiparticles or domain walls (see also Refs.~\cite{wurtz_variational_2020, wurtz_emergent_2020}), constructing effective Hamiltonians via the Schrieffer-Wolff transformation, numerically computing approximate eigenstates using efficient numerical methods such as the DMRG-X algorithm~\cite{khemani_dmrgx_2017}, designing optimal paths for quantum annealing protocols, suppressing dissipative losses in thermal machines and more. Interestingly, finding these optimal paths does not require diagonalizing the Hamiltonian of the system either exactly or approximately and can be done even in the thermodynamic limit.

We found that these optimal paths always start/terminate at the points corresponding to Hamiltonians exhibiting macroscopic degeneracies of the spectrum, which play a role similar to the role of quantum critical points in equilibrium phase diagrams. As we approach these singularities, the anisotropy between the optimal and the orthogonal directions diverges. The most divergent contributions to the adiabatic gauge potential are local and can be singled out either perturbatively or variationally. Increasing the support of the variational gauge potential, additional (weaker) divergences start to emerge, strongly affecting the flow diagram in their vicinity. Close to these singularities we can identify special dark states: mutual eigenstates of the Hamiltonian at the singular point and the divergent part of the adiabatic gauge potential. These dark states are highly robust against various time-dependent perturbations and can be efficiently locally dressed by the non-divergent part of the VAGP. They persist deep in the ergodic regime extending far away from the singularities. Physically, these dressed dark states correspond to spin configurations that can remain non-thermal for extremely long times. Our method provides a general prescription of finding such non-thermal states in interacting systems. 

Finally, we showed that the optimal directions are associated with the existence of local nearly-conserved operators. Thus there is an interesting and direct connection between our ability to perform efficient local adiabatic transformations along particular directions and the existence of long-lived operators, which are locally dressed deformations of the Hamiltonian along these optimal directions.

\begin{acknowledgements}
The authors thank Dries Sels, Maksym Serbyn, and Jonathan Wurtz for useful discussions and valuable comments. 
S.S. was supported by JSPS Overseas Research Fellowships (201860254). 
P.W.C. gratefully acknowledges support from a Francqui Foundation Fellowship from the Belgian American Educational Foundation (BAEF), Boston University's Condensed Matter Theory Visitors program, and EPSRC Grant No. EP/P034616/1. A.D. was supported by a grant of the Russian Science Foundation (Project No.~17-12-01587). A.P. was supported by the NSF Grant DMR-1813499 and the AFOSR Grant FA9550-16- 1-03. This research was supported in part by the International Centre for Theoretical Sciences (ICTS) during a visit for participating in the program - Thermalization, Many body localization and Hydrodynamics
(Code: ICTS/hydrodynamics2019/11).
\end{acknowledgements}

\onecolumngrid
\newpage
\appendix
\section{Derivation of the optimal direction}
\label{app:derive_opt_dir}

Following the notation of the main text, the derivative of the Hamiltonian with respect to the magnetic field $\vec \lambda=(h,g)\equiv \lambda (\cos\varphi, \sin\varphi)$ is $\partial_{\lambda}H = \cos\varphi Z + \sin \varphi X$. Denoting $\mathcal{A}_h$ and $\mathcal{A}_g$ as the adiabatic gauge potentials at $\varphi=0$ and $\varphi=\pi/2$ respectively and using the linearity of the full AGP we can write $\mathcal{A}_\varphi = \cos\varphi \mathcal{A}_h + \sin \varphi \mathcal{A}_g$. The squared norm of the AGP is given by
\begin{align}
\Tr \left[\mathcal{A}_\varphi^\dagger\mathcal{A}_\varphi\right]=\Tr\left[\mathcal{A}_h^\dagger \mathcal{A}_h\right] \cos^2\varphi + \Tr\left[\mathcal{A}_g^\dagger \mathcal{A}_g\right]\sin^2\varphi + \left(\Tr\left[\mathcal{A}_g^\dagger \mathcal{A}_h\right] + \Tr\left[\mathcal{A}_h^\dagger \mathcal{A}_g\right] \right)\cos\varphi\sin\varphi ,
\end{align}
which can be simplified to
\begin{align}
\Tr \left[\mathcal{A}_\varphi^\dagger\mathcal{A}_\varphi\right] = \Tr\left[\mathcal{A}_h^\dagger \mathcal{A}_h\right]\frac{1+\cos 2\varphi}{2}  + \Tr\left[\mathcal{A}_g^\dagger \mathcal{A}_g\right]\frac{1-\cos 2\varphi}{2}  + \left(\Tr\left[\mathcal{A}_g^\dagger \mathcal{A}_h\right] + \Tr\left[\mathcal{A}_h^\dagger \mathcal{A}_g\right] \right)\frac{\sin 2\varphi}{2}.
\end{align}
Differentiating this expression w.r.t. $\varphi$ and demanding this derivative to vanish then returns
\begin{align}\label{eq:app:derive_opt_dir}
\tan(2\varphi)  = \frac{\Tr\left[\mathcal{A}_g^\dagger \mathcal{A}_h\right] + \Tr\left[\mathcal{A}_h^\dagger \mathcal{A}_g\right]}{ \Tr\left[\mathcal{A}_h^\dagger \mathcal{A}_h\right]- \Tr\left[\mathcal{A}_g^\dagger \mathcal{A}_g\right]}.
\end{align}
This equation has two non-equivalent solutions (note that $\varphi$ and $\varphi+\pi$ are equivalent) corresponding to the minimum/maximum of the AGP norm. It is easy to see that for $\Tr\left[\mathcal{A}_h^\dagger \mathcal{A}_h\right] > \Tr\left[\mathcal{A}_g^\dagger \mathcal{A}_g\right]$ the minimum of the norm defining the optimal direction corresponds to the solution with $\varphi\in  \left[\frac{\pi}{4},\frac{3\pi}{4}\right]$ and the maximum to the solution in the interval $\varphi\in  \left[-\frac{\pi}{4},\frac{\pi}{4}\right]$. For $\Tr\left[\mathcal{A}_h^\dagger \mathcal{A}_h\right] > \Tr\left[\mathcal{A}_g^\dagger \mathcal{A}_g\right]$ the minimum and the maximum norm solutions are reversed. 

\section{Scaling of individual terms in the VAGP}
\label{app:scalingterms}
\begin{figure}[ht!]
	\includegraphics[width=1.\textwidth]{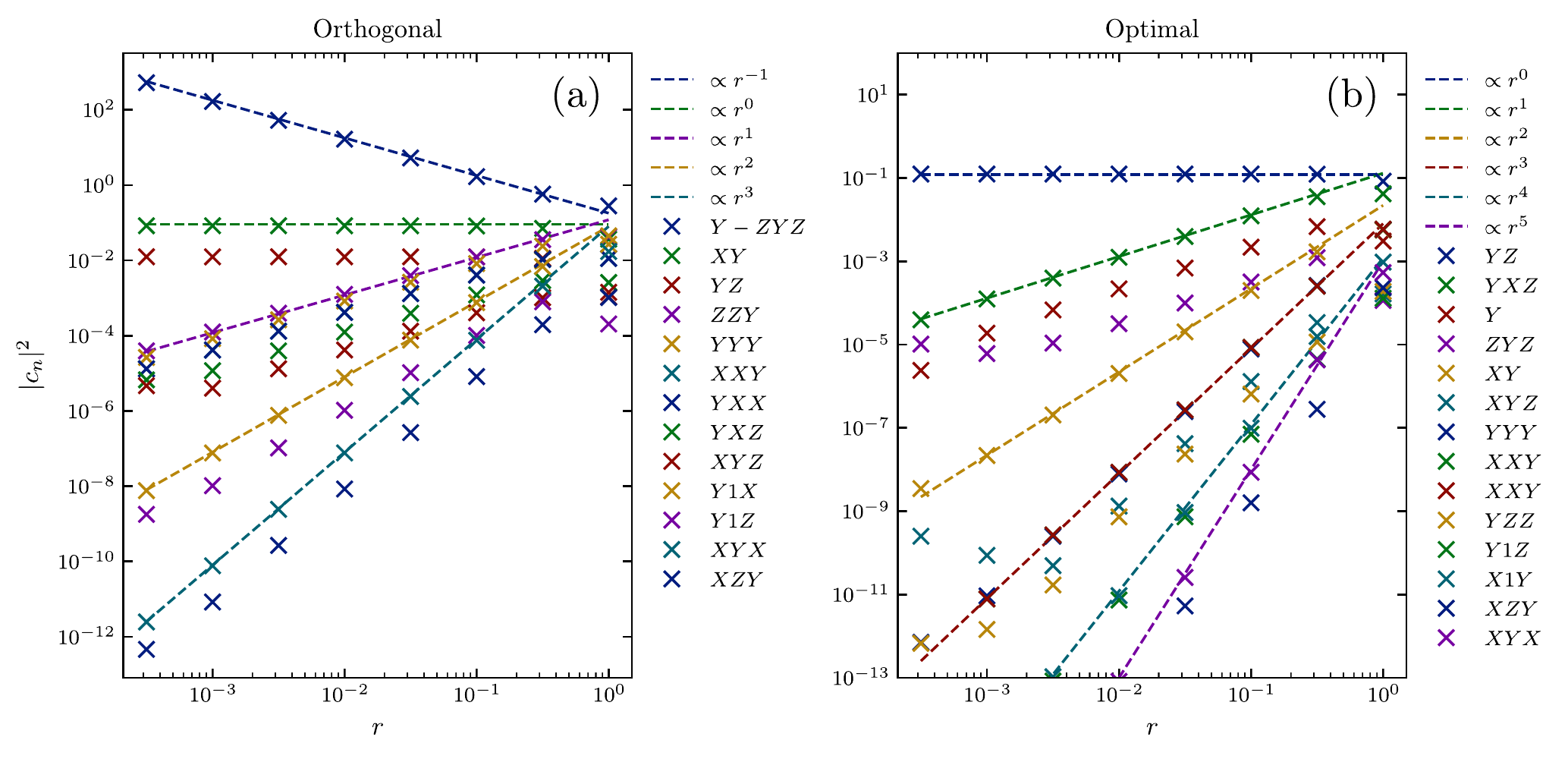}
	\caption{Scaling with $r$ of each term in the VAGP along the orthogonal (a) and optimal (b) direction. All the norms of the non-vanishing terms in the $3$-body ansatz are shown. The horizontal axis is $r=\sqrt{h^2+g^2}$ and the vertical one is the norm of each term $|c_n|^2$ (markers). In the orthogonal direction, the dominant term $O(1/r)$ is given by $Y-ZYZ$, whereas the dominant term along the optimal direction is $O(1)$ and given by $YZ+ZY$. All other terms exhibit higher-order scaling, where selected fits (full lines) are included as a guide to the eye. }
	\label{fig:scalingindterms}
\end{figure}

In this Appendix, we analyze the scaling of different operators appearing in the VAGP and compare those with the scaling predicted by the perturbative expansion. In Fig.~\ref{fig:scalingindterms}, we show the norms of the coefficients $c_n$ of the operator expansion of the VAGP (cf. Eq.~\eqref{variational ansatz}) within the 3-body variational ansatz near $(h,g)=(0,0)$. One can clearly observe the different power-law scalings of these coefficients with $r=\sqrt{h^2+g^2}$ and $g/h = 10$. Along the orthogonal direction [panel (a)] the $Y-ZYZ$  contribution diverges as $1/r$ for $r\to 0$, as expected from perturbation theory (cf. Eq.~\eqref{eq:agp_expansion}). The $r$-independent terms $XY$ and $YZ$ also agree with the perturbative calculations, and all remaining terms vanish at $r\to 0$ as various integer powers of $r$. In the optimal direction the divergent term is clearly absent and the remaining terms are similar to those in the orthogonal direction.

\section{State preparation in an orthogonal direction.}
\label{app:state preparation}

\begin{figure}[ht!]
	\includegraphics[width=.8\textwidth]{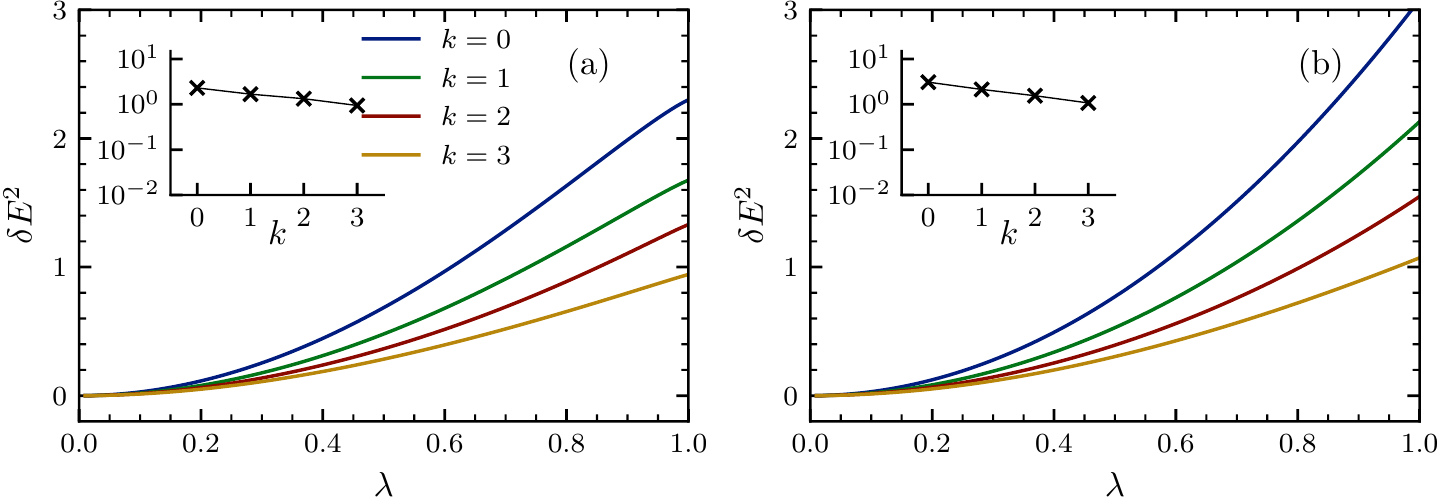}
	\caption{Same as Fig.~\ref{EneVar Optimal} for a path from $(\epsilon, 1-\epsilon)$ to $(0.5,0.5)$. Figure (a) corresponds to a finite protocol duration $T=2$ and (b) corresponds to infinitely-fast state preparation/dressing with $T=0$.}
	\label{fig:app:state preparation}
\end{figure}

In this Appendix we show the performance of the state preparation for the second orthogonal direction missing in Fig.~\ref{EneVar Optimal} (cf. Fig.~\ref{fig:Evardotlam}), for a path from $(\epsilon, 1-\epsilon)$ to $(0.5,0.5)$. The protocols are identical to those discussed in the main text. In the left panel of Fig.~\ref{fig:app:state preparation} we show the results for the unassisted protocol and the CD driving, and in the right panel we show the results for the infinitely fast (VAGP-only) protocol. In both cases the performance of the protocol is similar to that in the other orthogonal direction shown in panels (b) and (d) of Fig.~\ref{EneVar Optimal}.

\section{Non-symmetric dark/bright state preparation}

\begin{figure}[h]
	\includegraphics[width=0.8\columnwidth]{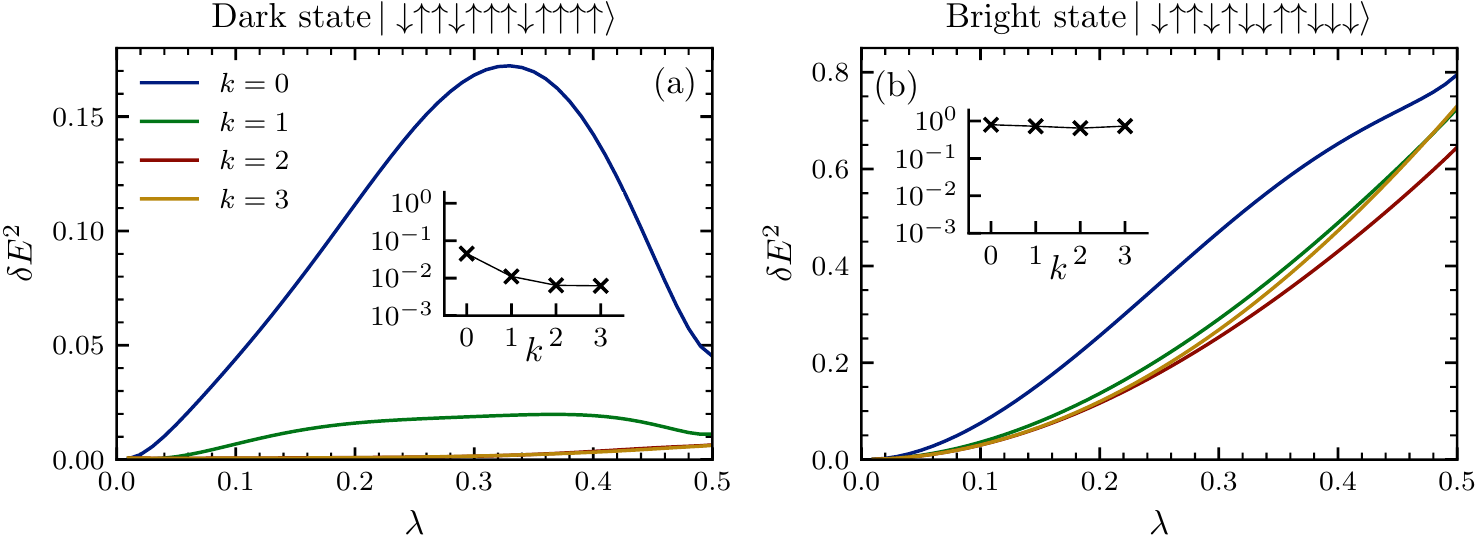}
    \caption{Same as Fig.~\ref{Dark states} with non-symmetric initial states. Panel (a) corresponds to a dark state $\state{\downarrow \uparrow \uparrow \downarrow \uparrow \uparrow \uparrow \downarrow \uparrow \uparrow \uparrow \uparrow }$, whereas panel (b) represents a random bright product state $=\state{\downarrow \uparrow\uparrow \downarrow \uparrow \downarrow \downarrow \uparrow \uparrow \downarrow \downarrow \downarrow  }$. Note again the different vertical scales in the two figures. Even for the unassisted protocol (blue lines), the energy variance is much smaller for the dark state compared to the bright state. Including local counterdiabatic terms further decreases the energy variance of the dark state and has only minimal effects on the bright state.}
    \label{Dark states 1}
\end{figure}

In Fig.~\ref{Dark states} we already compared the energy variance following a state preparation for particular symmetric dark/bright states using an unassisted protocol or by including a local variational CD term. As a reminder, the dark states are defined as the eigenstates of the Hamiltonian that are annihilated by the divergent part of the AGP, which near the $(2,0)$ singularity is proportional to $PYP$. It is easy to see that the dark states are those where $|\downarrow\rangle$ spins or pairs of such spins $|\downarrow\downarrow\rangle$ are separated from each other by at least two $|\uparrow\rangle$ spins next to each other. The bright states are those that violate this constraint. In Fig.~\ref{Dark states 1} we compare the performance of a randomly-chosen non-symmetric dark (a) and bright (b) states (see caption for details). The protocol is the same as in Fig.~\ref{Dark states}. It is clear that the results are similar to those shown in the main text for the symmetric dark/bright states.

\label{app:darkstates}

\section{Singularities of the AGP through degenerate perturbation theory}

In the Appendix, we present a short argument for why the radial direction is generally the optimal one near singular points. Let us assume that the Hamiltonian of the system can be written as
\begin{align}
	H=H_0 + \varepsilon V	,
\end{align}
where $H_0$ is a Hamiltonian describing a macroscopically-degenerate point and $V$ is some generic perturbation. The exact AGP can be represented in the degenerate eigenbasis of the instantaneous Hamiltonian as \cite{kolodrubetz_geometry_2017}
\begin{align}
	\mathcal{A}_\lambda =i {\sum_{n\neq m}} \state{{m}} {\bstate{{m}} \partial_\lambda H \state{{n}} 
	\over
	{\epsilon}_n - {\epsilon}_m,
	}
	\bstate{{n}}. \label{GP exact}
\end{align}
where $H \state{{n}} = {\epsilon}_n \state{{n}}$. It is clear from this expression that the AGP generally diverges in the limit $\epsilon\to 0$ as $1/\varepsilon$, since in first-order perturbation theory $\epsilon_n-\epsilon_m\approx \langle n | V| n\rangle- \langle m | V| m\rangle$. Note that in the proper eigenstates the matrix $V$ is approximately diagonal within each degenerate manifold. This divergence is, however, canceled if the deformation $\partial_\lambda H$ is diagonal (to the first order) in $\varepsilon$ in this basis. In particular, this is the case when $\lambda=\varepsilon$, which precisely corresponds to radial deformations defining the optimal directions (cf.  Sec.~\ref{Sec Anisotropy}).

\section{Details of the derivation of the perturbative AGP}
\label{app:perturbative}

In Eq.~\eqref{eq:X_t_def} we defined the $X$-magnetization operator in the interaction picture as $X(t) = e^{iH_0 t} X e^{-i H_0t}$ with $H_0 = ZZ+hZ$, through which one can express the expansion of the AGP in powers of the transverse field $g$. This operator satisfies the following equation of motion: 
\begin{align}
\frac{\partial}{\partial t}{X(t)}=i[H_0,X],\qquad X(0)=X.
\end{align}
It is easy to verify that there is a closed-form solution to this equation, reading:
\begin{align}
\label{eq:X(t)_Ising}
X(t)=&{X\over 4} \left(\cos(2(h+2)t)+2\cos(2h t)+\cos(2(h-2)t)\right) - {Y\over 4}\left(\sin (2 (h+2) t)+2\sin(2ht)+\sin(2(h-2)t)\right) \nonumber \\
&+{XZ+ZX\over 4} \left(\cos(2(h+2)t)-\cos(2(h-2)t)\right)+{ZXZ\over 4}\left(\cos(2(h+2)t)-2\cos(2h t)+\cos(2(h-2)t)\right) \nonumber\\
&-{YZ+ZY\over 4} \left(\sin(2(h+2)t)-\sin(2(h-2)t)\right) -{ZYZ\over 4}\left(\sin(2(h+2)t)-2\sin(2h t)+\sin(2(h-2)t)\right).
\end{align}

For the second-order contribution to the AGP we also need to compute $\chi(t)=\int_0^t d\tau X(\tau)$ (cf. Eq.~\eqref{eq:X_t_def}), which can be readily done as
\begin{align}
\chi(t) =& \int_0^t d\tau X(\tau) \nonumber\\
=&{X\over 8} \left({\sin(2(h+2)t)\over h+2}+{2\sin(2h t)\over h}+{\sin(2(h-2)t)\over h-2}\right) +{XZ+ZX\over 8} \left({\sin(2(h+2)t)\over h+2}-{\sin(2(h-2)t)\over h-2}\right) \nonumber\\
&+{ZXZ\over 8 }\left({\sin(2(h+2)t)\over h+2}-{2\sin(2h t)\over h}+{\sin(2(h-2)t)\over h-2}\right) \nonumber\\
&-{Y\over 8}\left({1-\cos (2 (h+2) t)\over h+2}+{2(1-\cos(2ht))\over h}+{1-\cos(2(h-2)t)\over h-2}\right) \nonumber\\
&-{YZ+ZY\over 8} \left({1-\cos(2(h+2)t)\over h+2}-{1-\cos(2(h-2)t)\over h-2}\right) \nonumber\\
&-{ZYZ\over 8}\left({1-\cos(2(h+2)t)\over h+2}-{2(1-\cos(2h t))\over h}+{1-\cos(2(h-2)t)\over h-2}\right).
\end{align}
Its commutators with $X(t)$ and $Z$ follow as
\begin{align}
&i [\chi(t), X(t)] = {1\over 4}(XY+YX)\left[{4\sin(4 ht)\over h^2-4}+{\sin(2(h-2) t)\over h+2}-{\sin(2(h+2) t)\over h-2}\right]\nonumber\\
&-{1\over 4} (YXZ+ZYX)\bigg[{2\sin(4 t)\over h^2-4}+{\sin(4(h+1) t)\over h(h+2)}-{\sin(4(h-1) t)\over h(h-2)}\biggr]\nonumber\\
&+{1\over 4}(XYZ+ZYX)\biggr[{2h \sin(2 h t)-2 \sin(4t)\over h^2-4} +{\sin(4(1+h) t)\over h(h+2)}-{\sin(4(h-1) t)\over h(h-2)}-{\sin(2(2+h)t)\over h}-{\sin(2(h-2)t)\over h}\bigg]\nonumber\\
&-{1\over 4}(ZXYZ+ZYXZ)\bigg[{2 \sin(2ht)+2\sin(4ht)\over h^2-4}-{\sin( 4(h-1) t)\over h(h-2) }-{\sin( 4(h+1) t)\over h(h+2) }-{\sin( 2(h-2) t)\over h(h+2) }-{\sin( 2(h+2) t)\over h(h-2) }\bigg]\nonumber\\
&+\text{[even terms in $Y$]},
\end{align}
and
\begin{multline}
i [\chi(t), Z]={1\over 4} Y \left({\sin(2(h+2)t)\over h+2}+{2\sin(2h t)\over h}+{\sin(2(h-2)t)\over h-2}\right)+
{1\over 4} (YZ+ZY) \left({\sin(2(h+2)t)\over h+2}-{\sin(2(h-2)t)\over h-2}\right)\nonumber\\
+{1\over 4} ZYZ\left({\sin(2(h+2)t)\over h+2}-{2\sin(2h t)\over h}+{\sin(2(h-2)t)\over h-2}\right)+\text{[even terms in $Y$]}
\end{multline}
Note that the even terms in $Y$ do not contribute to the AGP: as follows from e.g. Eq.~\eqref{GP exact} the AGP is explicitly imaginary for a real Hamiltonian (see also Ref.~\cite{sels_minimizing_2017}). These terms, however, will contribute to higher order corrections to the AGP.

Using these expressions together with Eqs.~\eqref{eq:A_Interaction_0} we can immediately recover the leading-order contributions to the AGP shown in Sec.~\ref{sec:perturbative}. Likewise, using~\eqref{eq:A_Interaction_1} we can find the first subleading corrections, which we will show below for different values of $h$:

\begin{itemize}

\item $h\neq 0,1,2$

\begin{align}
\mathcal A_g^{(1)}=& {1\over 4 h (4-h^2)}(XY+YX)-{1\over 8 (1-h^2)} (YXZ+ZXY-XYZ-ZYX) \nonumber\\
&\qquad+{3\over 8 h (1-h^2)(4-h^2)}(ZXYZ+ZYXZ) \\
\mathcal A_h^{(1)}=&-{h^4-2h^2+8\over 2 h^2 (h^2-4)^2} Y+{h\over (h^2-4)^2}(YZ+ZY)-{3h^2-4\over h^2 (h^2-4)^2}ZYZ
\label{eq:agp_pert_01}
\end{align}

\item{$h=0$}

\begin{align}
\mathcal A_g^{(1)}&= {1\over 8} (XYZ+ZYX-YXZ-ZXY) \\
\mathcal A_h^{(1)}&= -{1\over 4 h^2} (Y-ZYZ)
\end{align}

\item{$h=1$}

\begin{align}
\mathcal A_g^{(1)}&= {1\over 12}(XY+YX)-{1\over 32} (YXZ+ZXY)-{5\over 96}(XYZ+ZYX)-{5\over 96}(ZXYZ+ZYXZ) \\
\mathcal A_h^{(1)}&= -{7\over 18} Y+{1\over 9} (YZ+ZY)+{1\over 9} ZYZ
\end{align}

\item{$h=2$}

\begin{align}
\mathcal A_g^{(1)}&= {1\over 8}(XY+YX)+{1\over 24} (YXZ+ZXY)-{1\over 96}(XYZ+ZYX)-{1\over 12}(ZXYZ+ZYXZ) \\
\mathcal A_h^{(1)}&= -{1\over 8 (h-2)^2} (Y-ZY-YZ+ZYZ)=-{1\over 8 (h-2)^2} PYP
\end{align}

\end{itemize}

From the expansion~\eqref{eq:agp_pert_01} we recover both the singularities close $h=0$ and $h=2$, as discussed in the main text, as well as the emergence of a new singularity close to $h=1$, which one can check is proportional to $P(XY+YX)P$. One can further check that a similarly divergent term, also proportional to $P(XY+YX)P$, appears in the second order correction to $\mathcal A_h$, which we only show for completeness away from singularities, i.e. $h\neq 0,1,2$: 
\begin{multline}
\mathcal A_h^{(2)}={10 h^6 -5 h^4-35 h^2 +12\over 16 h^2 (h^2-1)^2 (h^2-4)^2}(XY+YX)-{h^6+12 h^4-30 h^2+8\over 8 h (h^2-1)^2 (h^2-4)^2}(XYZ+ZYX)\\
+{h\over 8 (h^2-1)^2}(YXZ+ZXY)+{23 h^4-61 h^2+20\over 16 h^2 (h^2-1)^2 (h^2-4)^2} (ZXYZ+ZYXZ).
\end{multline}
Collecting the terms that will be singular at $h=1$ appearing in the expressions for $\mathcal A_g^{(1)}$ and $\mathcal A_h^{(2)}$, we can obtain Eq.~\eqref{eq:Adiv_h1}.

\bibliography{VAGP_Apr28.bib}  

\end{document}